\documentclass[10pt,twocolumn,groupedaddress,pra,aps,showpacs,showkeys]{revtex4-1}
\usepackage{amsmath,mathrsfs,amsbsy,color,graphicx,bm,amsthm,amsfonts,dsfont,afterpage}
\usepackage[english]{babel}
\usepackage{tabularx,ragged2e}
\usepackage{soul}
\newcolumntype{C}{>{\Centering\arraybackslash}X}
\usepackage{amssymb,lipsum,attachfile}
\usepackage{amsmath,epsfig,epstopdf}
\usepackage{amsfonts,dsfont,amsthm,bibunits,mathpazo}
\usepackage{hyperref}
\usepackage{setspace}
\usepackage{multirow}

\newcommand{\be}{\begin{equation}}
\newcommand{\ee}{\end{equation}}
\newcommand{\bea}{\begin{eqnarray}}
\newcommand{\eea}{\end{eqnarray}}
\newcommand{\ket}{\rangle}
\newcommand{\bra}{\langle}

\begin{document}
\newtheorem{theorem}{Theorem}
\newtheorem{proposition}[theorem]{Proposition}
\newtheorem{corollary}[theorem]{Corollary}
\newtheorem{open problem}[theorem]{Open Problem}
\newtheorem{Definition}{Definition}
\newtheorem{remark}{Remark}
\newtheorem{example}{Example}

\title{Classes of topological qubits from low-dimensional quantum spin systems}
\author{Dong-Sheng Wang}
\affiliation{Institute for Quantum Computing and Department of Physics and Astronomy, \\University of Waterloo, Waterloo, Canada}

\begin{abstract}
  Topological phases of matter is a natural place for encoding robust
  qubits for quantum computation.
  In this work we extend the newly introduced class of qubits based on valence-bond solid models
  with SPT (symmetry-protected topological) order to more general cases.
  Furthermore, we define and compare various classes of topological qubits
  encoded in the bulk ground states of topological systems,
  including SSB (spontaneous symmetry-breaking),
  TOP (topological),
  SET (symmetry-enriched topological),
  SPT, and subsystem SPT classes.
  We focus on several features for qubits to be robust,
  including error sets, logical support, code distance and shape of logical gates.
  In particular, when a global U(1) symmetry is present and preserved,
  we find a twist operator that extracts the SPT order plays the role of a topological logical operator,
  which is suitable for global implementation.
\end{abstract}


\maketitle

\begin{spacing}{1.0}
\section{Introduction}

Qubits are the building blocks for quantum computers to be built
in the near future.
Qubits and quantum gates have been realized in many systems,
including superconductors,
trapped ions,
photons,
etc~\cite{NC00}.
Without passive protection, the coherence time is limited
and active quantum error correction is necessary~\cite{KL97,DiV00,Kni05}.
Physical encodings with a certain passive protection are also pursued,
such as the cat code in optical system~\cite{MLA+14},
and encodings via superconductors~\cite{BKP13}.
In the setting of quantum many-body system,
topological states provide promising candidate for robust qubits,
such as the well-known toric code~\cite{Kit03}.

In this work we define and compare various classes of topological qubits,
given the profound progress of topological phases of matter in recent years~\cite{ZCZ+19,Wen19s}.
Topological qubits are usually defined in purely topological systems,
while recently a class of topological qubits with symmetry protection is proposed~\cite{WAR18,Wang19}.
Therefore, it is important to compare the features of different classes of topological qubits.
We focus on qubits encoded in the bulk ground states manifold
of translationally invariant gapped systems,
although other kinds also exist,
such as encoding based on edge modes
or extrinsic defects~\cite{BLK+17}.
Phases of matter can be roughly classified according to
what the symmetry is,
whether and how the symmetry is broken or preserved.
In this work, the symmetry we consider can be
global, local or intermediate,
continuous or discrete.
The phases of matter we consider include gapped phases of
spontaneous symmetry-breaking (SSB) of a global symmetry,
purely topological (TOP),
symmetry-protection of a global symmetry (SPT) or
subsystem symmetry (SSPT).
The classification of TOP and SPT phases are relatively well understood
using group cohomology, cobodism, and tensor category~\cite{CGW11,SPC11,CGL+12,MR13,LKW17},
compared with the SSPT phases~\cite{YDB+18,Dev19,Gro19}.
Note that systems with subsystem symmetry,
such as `fracton' orders~\cite{VHF15},
also rely on geometry and play more intriguing roles in 3D spatial dimension.
In this work, we focus on 1D and 2D cases as the starting point
for the comparison of qubits.
For simplicity, we define the qubits considered as 'topological',
although there might be geometrical features.
As the ground-state degeneracy (GSD) required to encode a qubit can come from SSB or TOP order,
we define five classes of qubits that each rely on
SSB, TOP, SPT and SSB, SPT and TOP (also known as SET),
SSPT and SSB orders.
See Table~\ref{tab:qubits} for a brief summary of topological qubits
in dimension less than three we study
and their main features.

For encoding we use whole gapped phases instead of
a few states that are representative points in certain phases.
The encoding we consider is physical or on the hardware level, i.e.,
the encoded system is not only described by quantum states,
but also by other objects, such as Hamiltonian.
This is along the line of encoding a classical bit by the Ising model,
which does not require a fine-tuning of parameters such as temperature below the critical point,
see Fig.~\ref{fig:phases}.
As a result, the errors that should be considered
are not abstract or `digital',
instead they are analog errors that are allowed by the system.
For instance, if a global symmetry is preserved,
the natural errors should be symmetry-preserving, e.g.,
excitations of the system.

When there is a preserved global U(1) symmetry,
a twist operator can be defined and serves as a topological logical operator.
The twist can be understood in many ways,
and one of them is as a flux insertion inducing a topological geometric phase~\cite{Bru04},
which is protected by the gap.
The flux is a sum of local flux terms that can be nonuniform,
showing robustness against local perturbations.
We consider valence-bond solid (VBS) models with global SU($N$) symmetry
as a seminal class of SPT qubits, previously termed as VBS qubits for a special model~\cite{WAR18}.
This class of qubits can also support logical gates that are on higher levels of the Clifford hierarchy~\cite{NC00}.
The twist utilizes a U(1) subgroup of the global SU($N$) symmetry,
and can also be viewed as a way
to extract the SPT order of the system.
We use gapped phases for encoding
since the twist works differently for gapless phase~\cite{CYH97,SKK03,SKK04}
without a gap protection.

For a logical qubit, we find the two non-commuting logical operators
($\bar{X}$- and $\bar{Z}$- type rotations)
can be realized by topological or global operations.
For the SPT qubits,
$\bar{Z}$ is the global twist due to the SPT order,
$\bar{X}$ is the generator of the broken symmetry.
In the case of TOP qubits,
logical operators are all from Wilson loops~\cite{Wil74,Ogi81}.
Note here we call all loop operators as Wilson loops for simplicity.
Also Wilson loops can be viewed as the so-called `1-form' symmetry operators~\cite{KS14,GKS+15},
and TOP degeneracy can be viewed as the consequence of the SSB of 1-form symmetry~\cite{Wen19}.
Recall that a $q$-form symmetry acts on a $(D-q)$-dimensional manifold $\mathcal{M}^{D-q}$
in a system of spatial dimension $D$.
A global symmetry is 0-form regardless of $D$, for instance.
In the case of SET qubits,
a topological twist is equivalent to a corresponding Wilson loop operation.
For code properties, the Wilson loop determines the code distance, hence
plays more strict role than the twist.
However, the twist can be realized by external global fields, hence
benefits practical implementations without the need of local addressability.

In this work, we extend the construction of VBS qubits~\cite{WAR18},
which are SPT qubits with global continuous symmetry,
to more general SPT qubits that also allow global or subsystem discrete symmetry.
We compare features, mainly as quantum memory,
of SPT qubits with SET qubits and the more conventional TOP qubits,
treating all of them on the equal footing as candidates for
topological qubits and topological quantum computing.
Our study is model-based and certainly far from complete,
yet we believe our technique and the main features of these classes of qubits are generic.
This work is organized as follows.
In section~\ref{sec:pre},
encoding of topological qubits and the framework we employ are discussed.
In section~\ref{sec:vbs} the class of SPT qubits is introduced,
and the 1D VBS qubits, i.e., 1D SPT qubits with global U(1) symmetry, are studied in great details.
SPT qubits with discrete symmetry are deferred to Appendix~\ref{sec:app}.
In section~\ref{sec:2D} features of different classes of qubits
are analyzed and compared.
We conclude in section~\ref{sec:conc} with perspectives.

\begin{table}
  \centering
  \footnotesize{
  \begin{tabular}{|c|c|c|c|c|c|c|}
    \hline \hline
    qubits          & SSB         & TOP       & SPT          & SET         & SSPT                 \\ \hline
    Logical shape   & point,bulk  & string    & bulk         & bulk/string & string,bulk      \\ \hline
    Logical support & 1           & 1         & $L$          & 1           & $\sqrt{L}$      \\ \hline
    Logical distance& 1           & $\sqrt{L}$& $L$          & $\sqrt{L}$  & $\sqrt{L}$          \\ \hline
    Errors          & local       & TOP       & local/sym    & TOP/sym   & local/sym        \\ \hline \hline
  \end{tabular}}
  \caption{Table of classes of `topological' qubits we study
  encoded in degeneracy of ground subspace.
  For SSB, SPT, and SSPT qubits, the GSD come from SSB order.
  For TOP and SET qubits, the GSD come from TOP order.
  We will study the last three classes in more details since
  the SSB qubits, which are topologically trivial, and TOP qubits are well known.
  Here $L$ denotes the system size.
  For the specification of errors,
  `local' means the errors are local instead of being global,
  `TOP' ('sym') means the errors respect the topology (symmetry).
  Please refer to discussion in the main text for details.
  }\label{tab:qubits}
\end{table}

\section{Preliminary}
\label{sec:pre}

\subsection{Encoding of qubits}

Here we study primary defining features for robust qubits with topological protection.
We take a hybrid point of view as the combination of physical robustness and error-correction codes.
Namely, we take the qubits as a scheme of physical hardware encoding with natural error-resilience.
As the physical systems we consider are still abstract models instead of actual systems,
the well-known time scales, $T_1$ for relaxation and $T_2$ for dephasing,
usually involved as the character of a qubit,
are not studied in our setting.
Also we do not study active error-correction in depth as we currently focus on
the error-resilience from the system itself.

\begin{figure}
  \centering
  \includegraphics[width=.45\textwidth]{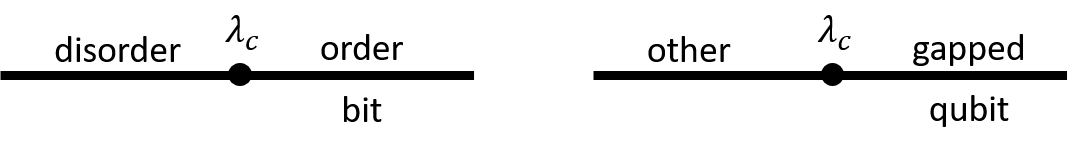}
  \caption{(Left) Phase diagram for 2D Ising model to encode a classical bit.
  The disordered phase is the high-temperature phase,
  and the ordered phase is the low-temperature phase that SSB a global $Z_2$ symmetry.
  The critical point $\lambda_c$ is a finite-temperature phase transition.
    (Right) Phase diagram used to encode a topological qubit.
    The `other' phase can be gapped or gapless.
    The qubit is encoded in a gapped phase.}
\label{fig:phases}
\end{figure}

Here we lay out our method for encoding of qubits.
Given a Hamiltonian $H(\vec{\lambda})$ of a model,
if in the parameter space of $\vec{\lambda}$
there is a gapped phase that breaks a certain symmetry,
then a logical qubit (or qudit) can be encoded in its bulk ground states.
The ground state degeneracy (GSD) determines the dimension of the logical space.
We do not require edge modes for encoding in particular.
This applies to the qubits that are studied in this work,
see Fig.~\ref{fig:phases},
including classical bits based on Ising models,
and qubits for models that show SPT or TOP orders.
A logical state is a state from the ground subspace,
and a logical basis is a set of orthonormal states that spans the qubit (or qudit) space,
in the large system-size limit if necessary.
A logical operator, $\bar{X}$ or $\bar{Z}$ for instance,
is defined to be the effective operation of an operator acting on logical states.
The same logical operator can be realized by many actual operators on the system.
The weight of a logical operator, or code distance of it,
is defined to be the minimal weight of those operators.
The weight of an operator is the size of the nontrivial support of it on the system.
Logical code distance, or logical distance for brevity,
is defined as the minima of the minimal weight of logical $\bar{X}$ and $\bar{Z}$,
or $\bar{X}$- and $\bar{Z}$- type rotations that are provided in the system.
We also use $d_x$ ($d_z$) to refer to the $\bar{X}$- ($\bar{Z}$-) type code distance.
Logical support
is defined as the minimal overlap of the supports of logical $\bar{X}$ and $\bar{Z}$,
or $\bar{X}$- and $\bar{Z}$- type rotations that are provided in the system.
Logical shape of a logical operator
is defined as the geometric support embedded in the system,
which could be a point, segment, string, or bulk etc.

Physical errors, or errors for short,
are operations that lead to excited states of $H$
or operations that are from the commutant of $H$.
These include the symmetry of $H$ and excitations of $H$, for instance,
and they are analog errors that are specific to a model $H$ and may not be arbitrary.
For instance, if a global symmetry is preserved, the natural errors should be symmetry-preserving.
A correctable set of analog errors $\{E_i\}$ are defined such that
the error-correction condition
$  P_\mathcal{C} E_i^\dagger E_j P_\mathcal{C}= C_{ij} P_\mathcal{C}$
is satisfied,
$P_\mathcal{C}$ is the projector to the code space,
and $\{C_{ij}\in \mathbb{C}\}$ form a hermitian matrix~\cite{KL97}.
For models that are not exactly solvable, such as valence-bond solids,
we will study approximate excitations that are well understood.

To define a good qubit, there are many other features studied in literature,
for instance, the disjointness~\cite{JKY18},
pieceable fault-tolerance~\cite{YTC16} of stabilizer codes,
and perturbative instability of the models~\cite{CL98,Gri04}.
In this work, we limited ourselves to the primary features listed above,
and we also leave fault-tolerant computation for future study.

In our framework of encoding,
a whole gapped phase is used for logical qubits.
A whole phase is used instead of a few representative points in it,
so the encoding is stable against perturbations of parameters $\vec{\lambda}$.
Furthermore, the employed phase needs to be gapped since
it can provide a certain symmetry-breaking leading to degeneracy required for encoding qubits,
along with well-defined logical gates.
The symmetries, either global or local, broken or preserved, determine the properties of the logical qubits,
and also put constraints on the errors that are natural for a system.
We also note that, although not study in the present work,
the symmetries also play central roles to realize quantum gates and computation.

\subsection{SSB qubit: Ising model}

To illustrate the framework followed in this work
we recall the scheme of encoding a robust classical bit in the
2D classical Ising model on a square lattice,
which is perhaps the most primary model for using spin systems as qubits.
To put it in the quantum setting, it is well known that
it can be viewed as the 1D quantum Ising model in a transversal magnetic field
\be H=-\sum_n X_n - \lambda \sum_n Z_n Z_{n+1}. \ee
There is a phase transition at $\lambda_c=1$,
and large (small) $\lambda$ corresponds to low (high) temperature
in the classical picture.
The critical point is identified by a notable self-duality of the system~\cite{FS78,Kog79}.
Define $\tilde{X}_n=Z_{n}Z_{n+1}$, $\tilde{Z}_n=\prod_{m<n}X_{m}$ on the dual lattice,
which satisfies the Pauli algebra
$\{\tilde{X}_n,\tilde{Z}_n\}=0$ and others.
The Ising model becomes
\be H=- \lambda \sum_n \tilde{X}_n -  \sum_n \tilde{Z}_n \tilde{Z}_{n+1}. \ee
It is clear that at $\lambda_c=1$ the model is self-dual,
and the order parameter $\bra \sum_n Z_n\ket=0$,
disorder parameter $\bra \sum_n \tilde{Z}_n\ket\neq0$
for the high-temperature phase ($\lambda<1$),
and $\bra \sum_n Z_n\ket\neq0$,
$\bra \sum_n \tilde{Z}_n\ket=0$
for the low-temperature phase ($\lambda>1$).

The low-temperature ordered phase is usually used to encode a classical bit,
see Fig.~\ref{fig:phases} (left).
This bit is said to be self-correcting, or thermally stable,
as it is protected by a finite-temperature phase transition from the disordered phase.
The ordered phase has SSB of a global $Z_2$ symmetry,
which provides the two-fold degeneracy for the whole spectrum.
The essential fact is that the encoding of the classical bit employs the whole ordered phase,
which not only includes ground states but also excited states with a finite total magnetization
in the classical picture.
The logical $\bar{X}$ is the generator of the broken $Z_2$ symmetry,
and can be realized by an external global magnetic field.
The logical $\bar{Z}$ (if treated as a qubit) is from the preserved 1-local symmetry,
which is a gauge symmetry, and $\bar{Z}=Z_n$ on any single site $n$.

When there is SSB for spatial dimension less than three,
the broken symmetry is finite due to Mermin-Wagner theorem,
and the generator of the broken symmetry will be one logical operator,
and its code distance will be linear with the system size.
We observe that there is a trade-off for the support of $\bar{X}$ and $\bar{Z}$
when there is a SSB of a global symmetry without other nontrivial symmetry preservation.
The support of $\bar{X}$, as the generator of the broken symmetry,
is the whole system,
while the support of $\bar{Z}$ is a constant.
Therefore, the logical support,
defined as the overlap for the support of $\bar{X}$ and $\bar{Z}$,
is a constant.
We will see qubits with logical support that scales with the system size later on.

There is no need for active error correction
as this can be simply done by lowing the temperature to keep the system in the ordered phase.
In the classical picture, the excitations are from local flips of spins,
hence the errors include local bit flips by $X_n$ and phase flips by $Z_n$.
The bit is robust against these local bit flips,
but not global ones which can lead to $\bar{X}$.
The bit is not robust against phase flips,
which is the reason for it being a good classical bit instead of a quantum one.
To design good qubits,
we have to employ more sophisticated models as studied in the following sections.

\section{1D VBS qubits}
\label{sec:vbs}

In this section we study symmetry-preserving orders and VBS qubits.
This is a generalization of the previous work on 1D code via SU($N$) VBS~\cite{WAR18},
where a topological twist operation and code properties were studied.
Here we broad our study to more general VBS,
and define VBS qubits and logical operations that are beyond the standard stabilizer codes.
We find VBS qubits behave quite differently from SPT qubits with discrete symmetries,
examples of which can be found in Appendix~\ref{sec:app}.
We first study the physics of twist operator in subsection~\ref{subsec:twist},
and then we analyze in details the spin-1/2 case,
the simplest VBS qubit, in subsection~\ref{subsec:dimer}.
We then generalize to higher-spin cases in subsection~\ref{subsec:1d_su2}
and $SU(N)$ cases in subsection~\ref{subsec:sun}.
Previously VBS and SPT models have been studied for the purpose of measurement-based
quantum computing~\cite{GE07,WAR11,Miy11,WSR17},
here our work demonstrate a different way of using them.
For 1D system, we require periodic boundary condition (PBC).
Also note that some authors may make a distinction between
valence-bond crystal and VBS,
which may be necessary in settings of condensed matter,
while here we find this is unnecessary and use VBS referring to
models of these kinds.

\subsection{Twist and logical operators}
\label{subsec:twist}

\begin{figure}[t!]
  \centering
  \includegraphics[width=.45\textwidth]{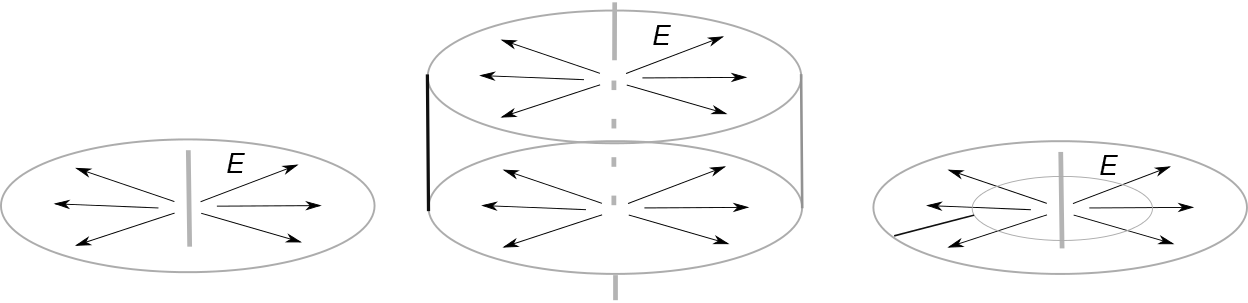}
  \caption{Implementation of twist by inserting a radial electric field.
  The 1D case (left) can be generalized to 2D case on a cylinder (middle) or a disc (right).
  The black line indicates the grouped sites treated as a single site for the twist operator.
   For the case of disc, however, the sites in a grouped site along a radial direction do not obtain the same phase due to the twist operator.
   }\label{fig:ac-twist}
\end{figure}

When there is a U(1) global symmetry that is respected by the ground state(s),
a twist operator along a periodic direction can be defined
\begin{equation}\label{eq:deftwist}
  F(\{\theta_n\},\theta):=\otimes_{n=1}^L e^{i\theta_n g_n},
 \end{equation}
 with
 \begin{equation}
  \sum_n (\theta_n-\theta_{n-1})=\theta,\;
  \theta_n-\theta_{n-1} \in O(1/L),\; \theta\in[0,2\pi],
\end{equation}
for $n$ as the site label,
$L$ as the length of the periodic direction.
The case $\theta=2\pi$ is called a full twist $F(\{\theta_n\},2\pi)$,
and the case $\theta_n-\theta_{n-1}=\Delta \theta, \forall n$
is called a uniform twist $F(\theta)$.
We will often use twist $F$ as the uniform full twist,
for which $\Delta \theta=2\pi/L:=\ell$.
The operator $g$ is the generator of the utilized $U(1)$ global symmetry,
e.g., we use $g=S^z$ for SU(2) case.
The order parameter $\lambda(G)$ of a ground state $|G\rangle$ is defined as
\begin{equation}\label{}
  \lambda(G):=\langle G| F(\{\theta_n\},2\pi)|G\rangle.
\end{equation}

It turns out there are multiple ways to look at the twist~(\ref{eq:deftwist}).
It is not hard to see that it is the exponent of a geometric phase $\Omega$
such that $e^{i \Omega}=\lambda(G)$ and
\begin{equation}\label{}
  \Omega=-i\oint_0^{2\pi} \langle G_\theta|\partial_\theta|G_\theta\rangle d\theta
\end{equation}
for $|G_\theta\rangle:=F(\{\theta_n\},\theta)|G\rangle$.
Due to the geometric phase interpretation,
the twist can be understood as a flux insertion,
hence the notation `F'.
For 1D system with PBC, there is only one twist.
This can be implemented by an external electric field for the SU(2) case~\cite{CYH97,SKK03,SKK04},
which is realized by putting a constant electrical charge $Q$ at the centre of the ring,
see Fig.~\ref{fig:ac-twist}.
Flipping the sign of the charge will flip the sign of the induced phase.
The electric field induces a global gauge transformation on the model.
Given the usual exchange interaction
\begin{equation}\label{}
  h=\vec{S}_r\cdot\vec{S}_{r+1}=
  (S_r^+S^-_{r+1}+S_r^-S^+_{r+1})/2+ S_r^zS^z_{r+1},
\end{equation}
the twisted exchange interaction
$h_\textsc{tw}(\theta):=F^\dagger(\theta) h F(\theta)$ and
\begin{equation}\label{eq:twistH}
  h_\textsc{tw}(\theta)=
  (e^{i\theta/L} S_r^+S^-_{r+1}+e^{-i\theta/L} S_r^-S^+_{r+1})/2+ S_r^zS^z_{r+1}
\end{equation}
contains a phase factor for each hopping term.
Furthermore,
the phase $\frac{\theta}{L}$ is the Aharonov-Casher (AC) phase~\cite{AC84},
which is the geometric phase of a magnetic moment $\vec{\mu}$
moving in electric field $\vec{E}$,  and
\begin{equation}\label{}
  \theta\propto L S \int_0^a d\vec{l}\cdot (\vec{e}^z\times \vec{E}(\vec{l}))
  =LSE(R)
\end{equation}
for spin $S$,
the unit vector $\vec{e}^z$ along z-direction orthogonal to the system plane,
and lattice spacing $a$.
The field $\vec{E}\propto \frac{Q}{r^2}$, and $E(R)$ is the field strength at the radius
$R=L/2\pi$.
We see that, for the full twist $\theta=2\pi$,
$Q \propto L/S$, which means in order to induce the desired twist operation,
the magnitude of charge is proportional to the system size while inversely to the magnitude of spin.
To make it size-independent,
an infinite line of charge can be used instead of a single charge,
and it is not hard to see the charge density $\rho\propto 1/S$.

Alternatively, the effect of the twist can be understood as a spin-orbit coupling
in the context of Dzyaloshinskii-Moriya (DM) interaction~\cite{Mor60}.
In general, DM term takes the form
\be H_\textsc{dm}=\sum_{ij} \vec{D}_{ij} \cdot (\vec{S}_i \times \vec{S}_j). \ee
To realize the twist, we only need the z-component of $D_{i,i+1}\propto \sin 2\pi/L$.
It turns out atomic quantum simulators using cold atoms in optical lattice~\cite{LSA12}
can realize the model~(\ref{eq:twistH}) as a special case.

In our encoding scheme,
the logical states are the singlet ground states of a certain VBS model.
They are not orthogonal if the system size is finite,
but become orthogonal when the thermodynamic limit is approached.
For SU(2) case, the broken symmetry is lattice translation,
and the encoding is for a single qubit.
For SU($N$) case the broken symmetry  can be lattice translation
or other symmetry such as parity,
and the encoding can be a qubit or qudit.
We find the logical $Z$-type rotation $\bar{Z}(\omega_N)$ is the twist operator $F$,
and the logical bit flip operator $\bar{X}$ is the generator of the broken symmetry.
When the broken symmetry is lattice translation $T$,
on a ground state $|G\rangle$ it holds
\begin{equation}\label{}
  TFT^\dagger |G\rangle= e^{i2\pi g_L} F|G\rangle,
\end{equation}
and $e^{i2\pi g_L}$ is proportional to identity up to a factor
as a certain order of $\omega_N$.
The two logical operators are Pauli $\bar{X}$ and $\bar{Z}$.
When the broken symmetry is parity $\Pi$ about a link,
which is complex conjugation for SU($N$) case,
it holds
\begin{equation}\label{}
  \Pi F\Pi^\dagger= F^\dagger.
\end{equation}
Due to the global SU($N$) symmetry,
it can be shown that $F^N \propto \mathds{1}$,
leading to the logical operators
\begin{equation}\label{eq:logic}
  \bar{X}=\begin{pmatrix} 0 & 1 \\ 1 & 0 \end{pmatrix},\;
  \bar{Z}(\omega_N):=\begin{pmatrix} \omega_N & 0 \\ 0 & \omega_N^* \end{pmatrix},\;
  \omega_N:=e^{i2\pi/N},
\end{equation}
and for each qubit the value of $N> 2$ is fixed.
For parity about a site, denoted by $R=T\Pi$,
it is also straightforward to obtain the logical operators above.
Some examples can be found in Table~\ref{tab:vbsqs}.

\subsection{Dimer-phase qubit}
\label{subsec:dimer}

Here we study a `dimer-phase' qubit for spin-1/2 VBS model in details.
For PBC with even number of sites $L$,
the Majumdar-Ghosh (MG) model~\cite{MG69}
\begin{equation}\label{}
  H_\textsc{mg}=\sum_{j=1}^L \vec{S}_j \cdot \vec{S}_{j+1} +
  \frac{1}{2}\vec{S}_j \cdot \vec{S}_{j+2}
\end{equation}
has two ground states $|\mathbf{L}\rangle$
and $|\mathbf{R}\rangle$, and
$|\mathbf{L}\rangle=|s\rangle^{\otimes L/2}$,
for singlet state $|s\rangle=\frac{1}{\sqrt{2}}(|01\rangle-|10\rangle)$
formed by two neighboring spins,
and $|\mathbf{R}\rangle$ is the same with $|\mathbf{L}\rangle$
after one lattice site translation.
A dimer is also known as a valence bond, or singlet.
The two ground states has SSB of the translation symmetry
of the Hamiltonian $H_\textsc{mg}$.

We define logical $|\bar{0}\rangle$ as $|\mathbf{L}\rangle$,
and $|\bar{1}\rangle$ as $|\mathbf{R}\rangle$.
The logical $\bar{X}$ is from the lattice translation by one site $T$,
or a Wilson loop:
firstly create a pair of spinons by breaking one singlet bond,
and then shift one of them around the system and annihilate the pair again.
In addition, it is easy to prove that
no local unitary operations of the form $U=\bigotimes_r U_r$ can serve as $\bar{X}$.
Instead, $\bar{X}$ can also be realized by a sequence of swap operations of local sites,
which cannot spread out or copy noises on a local site to more.

The gap protection of twist can be seen from
\begin{align}\label{}
  \langle \bar{0}|F(\theta)|\bar{0}\rangle &=
  (\cos\frac{\theta}{2L})^{L/2}\rightarrow 1, \\ \nonumber
  \langle \bar{1}|F(\theta)|\bar{1}\rangle &=
  (\cos\frac{\theta}{2L})^{L/2-1}\cos(\frac{\theta}{2}(1-\frac{1}{L})) \rightarrow \cos \frac{\theta}{2}.
\end{align}
As a result, for the full twist $\langle \bar{1}|F|\bar{1}\rangle=-1$,
and the minus sign comes from the fact that
$e^{i2\pi S^z}=-\mathds{1}$.
Also, for general value of $\theta\in (0,2\pi)$, $|\cos \frac{\theta}{2}|<1$,
which means the system will be excited,
and the value $-1$ from the full twist is protected by the gap of the system.

By grouping two sites, the two ground states of MG model belong to different SPT phases
protected by symmetry $SO(3)$~\cite{CGW11}.
For one ground state, say, $|\bar{0}\rangle$, the two spin-1/2 for each grouped site forms a singlet,
so there is no bond dimension and it belongs to the trivial phase of the cohomology
$H^2(SO(3),U(1))$.
For the other ground state $|\bar{1}\rangle$, however, there is one singlet bond between each two grouped sites,
so the bond dimension is two.
This state can be written as a matrix-product state (MPS)~\cite{Sch11}
with the set of matrices at each site as Pauli matrices $\{\sigma_i\}$
for on-site basis $|i\rangle$ formed by singlet and triplets.
The nontrivial twist factor $-1$ is the SPT index of $|\bar{1}\rangle$.
However, without grouping of sites the assignment of SPT index becomes relative.
Also it seems the two ground states are distinguishable locally, e.g.,
by identifying the two-local configuration to be singlet or triplet,
yet this breaks the global SU(2) symmetry.
Instead, the twist is the topological operation that respects the symmetry
and extracts the different SPT indices of them.

More generally, we consider
\begin{equation}\label{}
  H_\textsc{D}=\sum_{j=1}^L \vec{S}_j \cdot \vec{S}_{j+1} +
 J\vec{S}_j \cdot \vec{S}_{j+2} + B \sum_{j=1}^L S^z_j.
\end{equation}
$H_\textsc{mg}$ is a special case of $H_\textsc{D}$.
For variables $J$ and $B$ in a certain range
there is a dimer phase, i.e., ground states are product of singlet pairs
from nearest-neighbor sites~\cite{WA96}.
The essential properties we use are that
there is a double-degeneracy of the spectrum of $H_\textsc{D}$,
and there is a global U(1) symmetry (rotation along z-direction).
The double-degeneracy provides a two-dimensional space for our logical qubit,
and the global U(1) symmetry allows a flux insertion (twist) that plays the role of a logical operator.

Now we discuss robustness of qubit against noises and excitations.
The well-studied excitations are spinons, also known as solitons~\cite{SRF+08}.
The solitons form domain walls between the two ground states.
The solitons are deconfined as they can move without causing a net energy cost.
Bit-flip type errors will be spinon drift,
and they are correctable except the Wilson loop.
The logical $\bar{X}$ is a global operation
so it is not straightforward to mimic by thermal noises.
Furthermore, an energy barrier for $\bar{X}$ can also be introduced
from the Hamiltonian via spin-Peierls mechanism~\cite{HTU93},
which may due to spin-phonon interactions
and usually introduces the staggered terms $(1-\delta(-1)^r) \vec{S}_r \cdot \vec{S}_{r+1}$,
which explicitly breaks the lattice translation symmetry by odd number of sites,
and induces a confining force between spinon pairs.

The twist $F$ is based on a geometric phase.
In our case, the geometric phase is topological
as it is proportional to the winding number around the spin ring.
Due to the topological feature of $F$, it is robust against thermal noise, i.e.,
it is hard for the thermal noise to induce a logical $\bar{Z}$ operation.
For random unitary $U=\otimes_n U_n$ with $U_n\in SU(2)$,
the full-twist condition and the global symmetry are violated,
and bonds are actually broken and the system is destroyed.
So we shall only consider noises that almost respect the global symmetry.
Now for random unitary $U=\otimes_n R^z_n$ with $R^z_n\in SU(2)$ and close to identity,
hence almost respect the global symmetry,
the full-twist condition ($\theta=2\pi$) is violated in general.
This close-to-identity error (noise) is correctable as it has null action on the code subspace.

At low temperature $\beta_\text{th}$,
the solitons will on average distribute evenly on the system
with density $\rho\propto L e^{-\beta_\text{th}\Delta}$ for $\Delta$ as the gap of a single soliton.
The total magnetization $M$ is a thermal average $M=\sum_i p_i m_i$
for $p_i=\langle \psi_i|\rho_\text{th}|\psi_i\rangle$ as Boltzmann distribution,
$m_i$ as the magnetization of eigenstate $|\psi_i\rangle$.
For eigenstate $|e\rangle$, if its magnetization is $m$,
$\langle e|e^{-i\ell \sum_n S^z_n}|e\rangle=e^{i\ell m}\simeq 1$ if $m\ll L$.
We verify that, for lattice translation $T$ and the twist $F$,
\begin{equation}\label{eq:logicmg}
  \{T, F\}=0,\; T^2=F^2=\mathds{1},
\end{equation}
holds for the low-lying spectrum and the whole dimer phase~\cite{Aff88,Bon89}.
The relation~(\ref{eq:logicmg}) is the basic for taking a whole dimer phase
at finite low temperatures as a stable logical qubit.
This means we can encode logical qubit into low-lying spectrum,
and the whole Hilbert space has a junk part
\begin{equation}\label{}
  \mathcal{H}\ominus \mathcal{K}= \mathcal{H}_0\oplus \mathcal{H}_1.
\end{equation}
For a fixed number of density of solitons,
there is a trade-off between the system size and the value of temperature.
Proper choices of them would affect the practical performance of a qubit.
At low temperature, no active error correction is required as the Hamiltonian itself
provides a passive protection of the qubit against noises,
although the lifetime of the qubit is still finite.
This also encourages an effective description of the qubit by quantum field theory.

It is well developed that the long-wavelength behavior of 1D Heisenberg spin chain
close to phase transition is described by a Wess–Zumino–Witten model~\cite{AH87,Gia04,GNT04,Tsv07},
which is further equivalent to a sine-Gordon model~\cite{Sch86,CPV98}.
To describe the qubit, we start with a simple sine-Gordon Hamiltonian
\begin{equation}\label{}
  H_\text{sG}= \int dx \frac{1}{2}[ \Pi^2 + (\partial_x \phi)^2 ] + g \cos \beta \phi
\end{equation}
with variables $g$ and $\beta$.
Note here $\phi, \Pi$ are field operators, and they are $\phi(x), \Pi(x)$ in full notation and satisfy
$[\phi(x), \Pi(y)]=i\delta(x-y)$.
The dual field of $\phi$ is $\theta$ such that $\Pi= \partial_x \theta$.
The scaling dimension of the nonlinear term $\cos \beta \phi$ is $d= \frac{\beta^2}{4\pi}$,
which, in the sense of renormalization group,
is relevant if $d<2$, irrelevant if $d>2$, and marginal if $d=2$.
The model $H_\text{sG}$ depends on $\beta$ significantly~\cite{GNT04,Tsv07}:
it is gapped for $\beta^2< 8\pi$, and becomes gapless otherwise.
The gapped phase is dimerized due to breaking of lattice translation symmetry.

To make a connection with the original spin picture,
we note that
\be S^z_x \propto c_1 \partial_x \phi + c_2 (-1)^x \sin \beta\phi/2 \ee
for constants $c_1$, $c_2$.
The dimer order parameter is the real part of the vertex operator $z:= e^{i\beta\phi/2}$,
which is pinned to the value $\pm 1$ for the two dimer ground states.
The field $\phi$ is periodic $\phi=\phi+ 4\pi n/\beta $ for integer $n$,
and the lattice translation by one site $T$ acts as $T: \phi \rightarrow \phi + 2\pi/\beta,$
which flips the sign of the dimer order parameter.
A staggering $\delta$ can induce an additional term $\cos \beta \phi/2$ for explicit dimerization,
while the 2nd-nearest neighboring interaction affects the term $\cos \beta \phi$
for spontaneous dimerization~\cite{EA92,WA96,GNT04}.
The explicit dimerization induces the confinement of spinons to form bound states,
which will reduce the probability of the logical $\bar{X}$ error for the qubit.

\begin{figure}
  \centering
  \includegraphics[width=.2\textwidth]{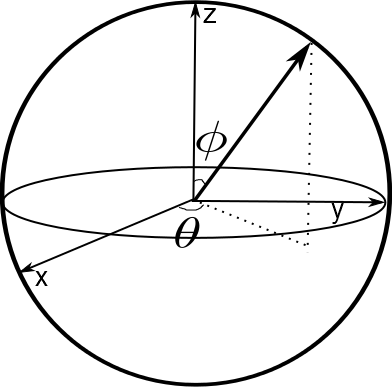}
  \caption{The Bloch sphere and Bloch vector to represent the logical qubit state.
  $(\phi, \theta)$ are the field operators in sine-Gordon field theory.
  The low-lying states of the dimer phase correspond to the Bloch vector
  pointing towards the positive or negative z-direction.}\label{fig:bloch}
\end{figure}

For the gapless phase, the vertex operator can be interpreted as an operator
that pumps a fermion between the two brunch of the Fermion surface,
and it is known that the value of $z=0$.
This is similar to the twist operator, which will create a spin-wave excitation.
For the gapped phase, due to the exponentially decaying spin correlation function
the twist operator does not create a spin wave,
instead the twist will induce a nontrivial action (logical $\bar{Z}$) on the ground subspace.
Again this is similar to the vertex operator which is pinned to fixed values
for the two ground states.
Indeed, it has been argued that the twist operator is equivalent to the vertex operator $z$~\cite{NV02,NT02}.
Therefore, we identify
$F\equiv e^{i\beta \phi/2}$ and $TFT^\dagger =-F$ holds for low-lying spectrum.
This is just the two logical operators $\bar{X}=T$, $\bar{Z}=F$ defined before.
From sine-Gordon model, we can use $(\phi, \theta)$ as the polar coordinates of the Bloch vector
of the logical qubit, see Fig.~\ref{fig:bloch}.
Note here $\phi$ and $\theta$ are field operators instead of c-numbers.
The ground states of the dimer phases (related by translation $T$)
correspond to the Bloch vector pointing to the z direction or its negative.
This can be viewed as a spontaneous breaking of $Z_2$ symmetry,
which is the lattice translation by one site.

\subsection{1D $SU(2)$ VBS qubits}
\label{subsec:1d_su2}

In this section we study 1D $SU(2)$ VBS model for general spin values $S$,
which could be half integers or half odd-integers.
We find the 1D parent Hamiltonian~\cite{MG69,Kle82,AKLT88} could include three-local interactions,
and its ground states may break translation symmetry by one lattice site.
For the encoded qubit,
we find the $\bar{X}$ is the generator of the broken translation symmetry,
and $\bar{Z}$ is the twist.
Different from the spin-1/2 case,
the ground states are more complicated,
and there are more types of excitations.
Despite this, the universal features can be characterized by effective field theories,
similar with the spin-1/2 case.
The field theories also apply to the $SU(N)$ cases,
while we defer this for a separate study.
We find one of the merits of higher-spin VBS qubits is that
spinons can be confined, hence reducing the probability of making logical errors.

A valence bond is a $SU(2)$ singlet formed by two spin-1/2.
We denote a valence bond state as $|\Xi_{mn}\rangle$ for $m+n=2S$,
and integers $m,n\geq 0$~\cite{GKM90,NT02}.
The on-site spin-$S$ is a projection from $2S$ spin-1/2,
and there are $m$ ($n$) bonds to the left (right) of this site.
Alternatively, $m$ bonds can be treated as a single bond of two spin-$m/2$.
For PBC there are even number of sites.

\begin{figure}[t!]
  \centering
  \includegraphics[width=.35\textwidth]{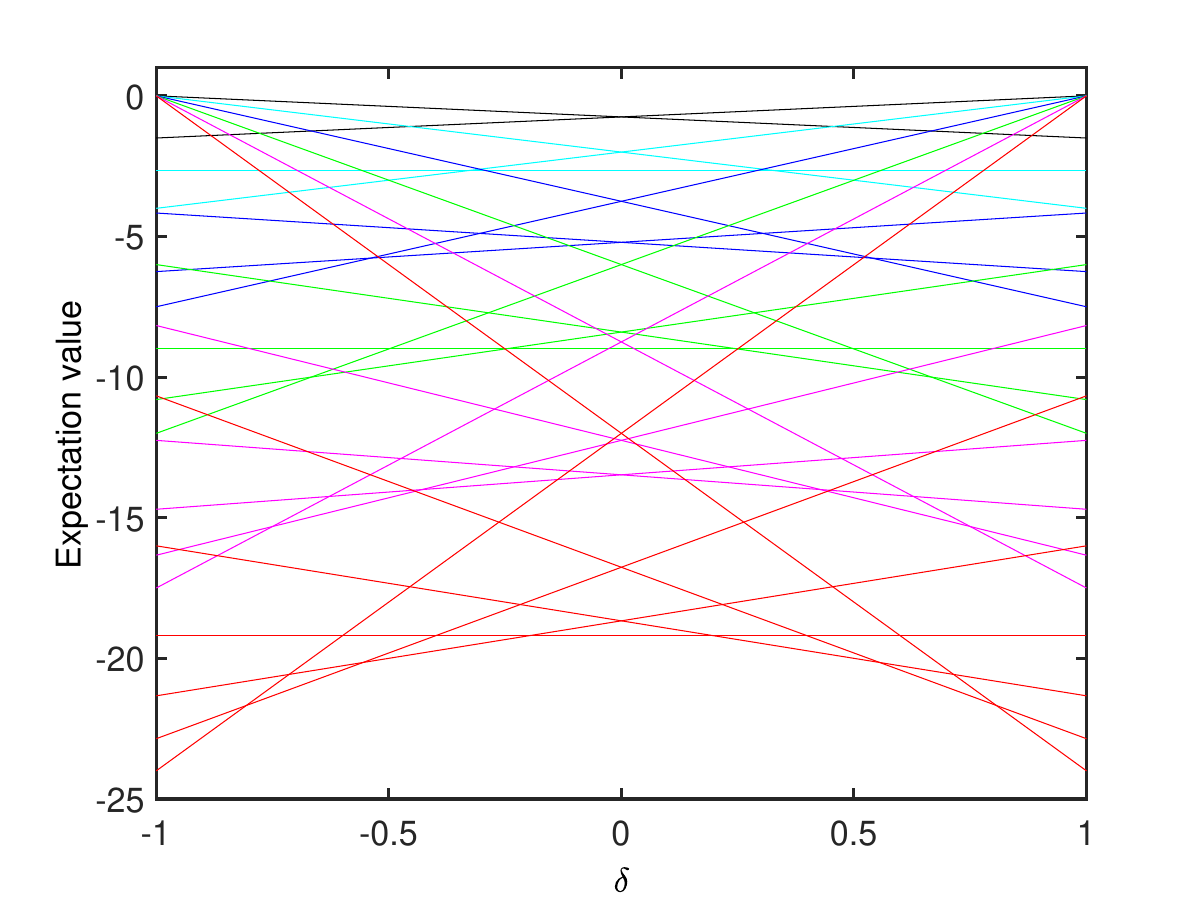}
  \caption{The expectation value of the sum of two neighboring interaction terms
  on VBS states $|\Xi_{mn}\rangle$ for different spins
  as a function of spin-Peierls staggering $\delta$.
  The spins are $S=1/2$ (black), $S=1$ (cyan), $S=3/2$ (blue),
  $S=2$ (green), $S=5/2$ (magenta), $S=3$ (red).}\label{fig:spin-Peierls}
\end{figure}

Each $|\Xi_{mn}\rangle$ as a MPS can be expressed as
\begin{equation}\label{eq:vbssu2}
  |\Xi_{mn}\rangle=\sum_{i_1,\dots, i_L} \text{tr}(A_{i_1}B_{i_2}\cdots A_{i_{L-1}}B_{i_L})|i_1\cdots i_L\rangle
\end{equation}
with two types of matrices
$A_i$ of size $(n+1)\times (m+1)$ and $B_i$ of size $(m+1)\times (n+1)$~\cite{TS95,NT02}.
We will also study a convenient fermion representation later on.
In our setting the MPS is also known as valence-bond states
due to the global continuous symmetry,
which provide a concise description for the generic features of
the Heisenberg interactions
\begin{equation}\label{}
  H=\sum_{r=1}^L \vec{S}_r \cdot \vec{S}_{r+1}.
\end{equation}
For instance, for spin-1 the ground states of Heisenberg model
and AKLT model are in the same phase, the well-known Haldane phase~\cite{Hal83,Hal83b,AKLT88}.
States $|\Xi_{mn}\rangle$ can be used to study phase transitions.
A common way to induce energy differences among $|\Xi_{mn}\rangle$ is the spin-Peierls effects,
as the spin-1/2 case.
We find there are generic features for the energy of states $|\Xi_{mn}\rangle$
as a function of staggering $\delta$.
In Fig.~\ref{fig:spin-Peierls} we plot the expectation value
of the sum of two neighboring interaction terms
on VBS states for spins up to 3.
We can see that the slope is proportional to $|m-n|$,
and there is no slope when $m=n$ as the staggering effects cancel out.
At $\delta=0$ for integer spin $S$,
the uniform VBS $|\Xi_{SS}\rangle$ always has lower value of energy than other states.

To take a certain states $|\Xi_{mn}\rangle$ as exact ground states,
parent Hamiltonian can be constructed from projectors~\cite{AKLT88}.
With the projector $P_{s'}(i,i+1,i+2)$ to spin $s'$ sectors on
three neighboring sites,
the following Hamiltonian
\begin{equation}\label{}
  H=\sum_i \sum_{s'=S+1}^{3S} P_{s'}(i,i+1,i+2)
\end{equation}
takes all states $|\Xi_{mn}\rangle$ as ground states,
i.e., the GSD is $2S+1$.
Furthermore,
a pair of VBS $|\Xi_{mn}\rangle$ and $|\Xi_{nm}\rangle$ can be selected out
to be the ground states of a parent Hamiltonian.
For the generic case, both two-local and three-local terms are needed.
Given $m, n$, w.l.o.g. let $m>n$,
then the projector $\sum_{s=m+1}^{m'}P_s(i,i+1)$ lifts up the energy
of states $|\Xi_{m'n'}\rangle$ and $|\Xi_{n'm'}\rangle$ for $m'>m$,
and the projector $\sum_{s=(m'-n')/2}^{(m-n)/2-1}P_s(i,i+1,i+2)$ lifts up the energy
of states $|\Xi_{m'n'}\rangle$ and $|\Xi_{n'm'}\rangle$ for $m'<m$.
Now we find the parent Hamiltonian of $|\Xi_{mn}\rangle$ and $|\Xi_{nm}\rangle$ takes the form
\begin{equation}\label{}
\begin{split}
  H=&\sum_i \sum_{s'=S+1}^{3S} P_{s'}(i,i+1,i+2) \\
  &+\sum_{s'=0}^{|m-n|/2-1} P_{s'}(i,i+1,i+2)\\
  &+\sum_{s'=\max(m,n)+1}^{2S} P_{s'}(i,i+1).
\end{split}
\end{equation}
The Hamiltonian can be simplified for special cases.
For integer spin $S$, only two-local interactions
\begin{equation}\label{}
  H=\sum_i \sum_{s'=S+1}^{2S} P_{s'}(i,i+1)
\end{equation}
are needed for the state $|\Xi_{SS}\rangle$ as the unique ground state
as it does not break the translation symmetry.
For the fully dimerized states $|\Xi_{2S,0}\rangle$ and $|\Xi_{0,2S}\rangle$,
its parent Hamiltonian is
\begin{equation}\label{}
  H=\sum_i P_{(S)}(i,i+1,i+2),
\end{equation}
and $P_{(S)}$ denotes the projection to all spin sectors except $S$.

We now identify the logical operators $\bar{X}$ and $\bar{Z}$.
It is not hard to see that in the large system size limit
\begin{equation}\label{}
 \langle \Xi_{mn}|F |\Xi_{mn}\rangle=(-1)^n,\;
 \langle \Xi_{nm}|F |\Xi_{nm}\rangle=(-1)^m.
\end{equation}
We will compute this via fermion representation later on.
For VBS with half-integer spins, we can always choose two VBS states with opposite SPT index
as degenerate ground states to encode a qubit.
From
\begin{equation}\label{}
  T F T^\dagger =F e^{-i\ell \sum_n S^z_n} e^{i2\pi S^z_1},
\end{equation}
for half-integer spin, $e^{i2\pi S^z_1}=-\mathds{1}$.
For phases with singlet ground states,
it holds $T F T^\dagger =-F$.
Now it is clear to see that the logical operators are
\begin{equation}\label{}
  \bar{Z}=F,\; \bar{X}=T.
\end{equation}
We know that $\bar{X}$ can be also be
done using Wilson loop, namely,
firstly create a pair of domain walls by breaking singlet bonds,
and then shift one of them around the system and annihilate the pair again.

For errors on the system, we can understand the generic properties from excitations.
Excitations of VBS models are usually described as solitons and pseudo-solitons~\cite{SRF+08}.
Solitons, which by definition separate ground states, are in general deconfined,
hence can induce logical bit flip error if they can move around the system,
forming a Wilson loop.
However, this process is very unlikely and suppressed by the system size
since a soliton can drift either towards or away from the other one,
or by the spin-Peierls effects.
Pseudo-solitons, which by definition separate a ground state from an excited state,
are in general confined, hence will not cause logical errors.
When pseudo-solitons dominate over solitons,
which require lowering the temperature and other mechanism,
the performance of a logical qubit would be better.

We remark on a difference between integer and half odd-integer spins,
which has been a notable point in spin systems~\cite{Hal83}.
Recall that $n+m=2S$.
Integer spin-$S$ is a linear rep of $SO(3)$.
For this case, if $n$ is odd,
then the virtual spin-$n/2$, and also spin-$m/2$, is a nontrivial projective rep
of $SO(3)$.
If $n$ is even,
then $m$ is also even and the order is trivial.
For half odd-integer spin $S$,
$n$ is odd if $m$ is even, and vice versa,
and they cannot both be even or odd.
That is to say, if $|\Xi_{mn}\rangle$ has trivial order,
then $|\Xi_{nm}\rangle$ has nontrivial order.
The on-site rep becomes linear if two sites are grouped together,
and the virtual rep is spin $n/2$ or $m/2$,
and it is clear only one of them is a nontrivial projective rep
of $SO(3)$.
This agrees with the phase classification by projective representation method~\cite{CGW11}.
For VBS models with integer spins,
the ground state can be unique without breaking the lattice translation symmetry.
Excited states or edge states have to be used to encode a qubit~\cite{GKS19,WOZ19},
which slightly deviates from the main encoding scheme using ground states,
so we do not study this further.

\subsection{1D SU($N$) VBS qubits}
\label{subsec:sun}

\begin{table*}
  \centering
  \caption{Examples of SU(3) and SU(4) 1D VBS qubits.
  In the table,
  `unique' means the ground state is unique,
  and an encoding with excited states or edge states is inevitable.
  `T' is a lattice translation operator,
  `$\Pi$' is a parity operator that exchanges irreps $\mathbf{N}\leftrightarrow \mathbf{\bar{N}}$,
  `R' is a reflection about a site and $R=T\Pi$.
  `H term' refers to the basic term in a parent Hamiltonian.
  $P^{n}_{(\lambda})$ refers to a projector onto all other irreps except $\lambda$ acting on $n$ neighboring sites,
  and $P^{n,A (B)}_{(\lambda)}$ refers to that when the first site is in sublattice A (B).
  The ``uc-size'' refers to the unit-cell size, which is the minimal number of sites such that the congruence class is [0].}
  \label{tab:vbsqs}
  \vspace{.2in}
  \begin{tabular}{|l|c|c|c|l|c|}
    \hline \hline
    System & Types & uc-size & Ground states & H term & Codes \\ \hline
    SU(3), $\mathbf{3}$  &  II & 3 &  \includegraphics[width=.1\textwidth]{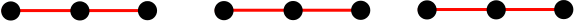}
    & $P^{\{4\}}_{(3,\bar{6})}$  & qutrit, $\langle \bar{X}=T, \bar{Z}=F\rangle$ \\ \hline
    SU(3), $\mathbf{6}$ &   II & 3 & \includegraphics[width=.1\textwidth]{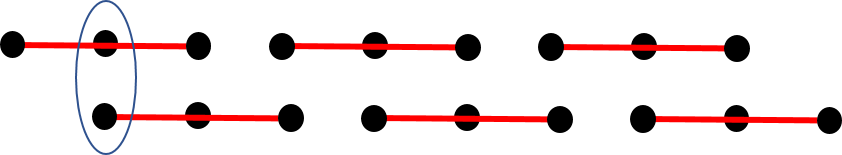}
    & $P^{\{4\}}_{(\bar{3},6,\bar{15})}$  & qutrit, $\langle \bar{X}=T, \bar{Z}=F\rangle$ \\ \hline
    SU(3), $\mathbf{10}$ & II & 1 &    \includegraphics[width=.07\textwidth]{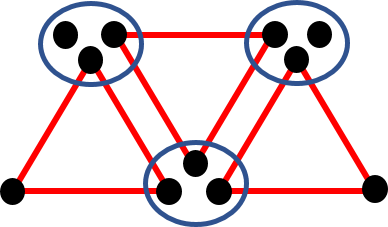}
    & $P^{\{2\}}_{(\bar{10},27)}$  & unique \\ \hline
    SU(3), $\mathbf{8}$ &    I & 1 &    \includegraphics[width=.08\textwidth]{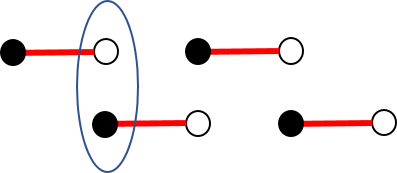}
    & $P^{\{2\}}_{(\bar{3}\otimes 3)}$  & qubit, $\langle \bar{X}=\Pi, \bar{Z}(\omega_3)=F\rangle$ \\ \hline
    SU(3), $\mathbf{27}$ &  I & 1 &    \includegraphics[width=.1\textwidth]{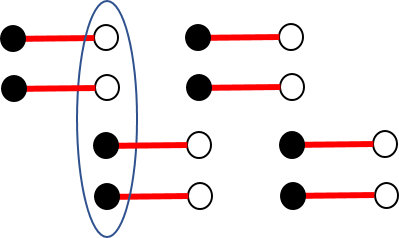}
    & $P^{\{2\}}_{(\bar{3}\otimes 3 \otimes \bar{3}\otimes 3)}$  & qubit, $\langle \bar{X}=\Pi, \bar{Z}(\omega_3)=F\rangle$ \\ \hline
    SU(3), ($\mathbf{3}$, $\mathbf{\bar{3}}$) &     III & 2 &     \includegraphics[width=.1\textwidth]{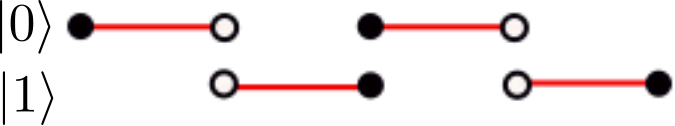}
    & $P^{\{3,A\}}_{(3)}$, $P^{\{3,B\}}_{(\bar{3})}$  & qubit, $\langle \bar{X}=R, \bar{Z}(\omega_3)=F\rangle$ \\ \hline
    SU(3), ($\mathbf{6}$, $\mathbf{\bar{6}}$) &     III & 2 &    \includegraphics[width=.1\textwidth]{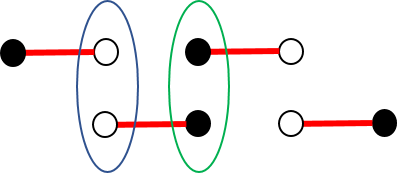}
    & $P^{\{2\}}_{(\bar{3}\otimes 3)}$  & unique \\ \hline
    SU(3), ($\mathbf{10}$, $\mathbf{\bar{10}}$) & III & 2 &    \includegraphics[width=.08\textwidth]{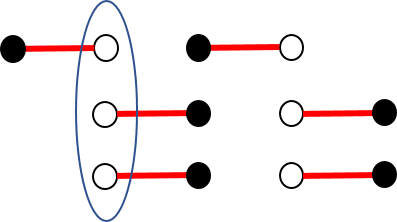}
    & $P^{\{3,A\}}_{(3\otimes 3\otimes 3)}$, $P^{\{3,B\}}_{(\bar{3}\otimes\bar{3}\otimes\bar{3})}$  & qubit, $\langle \bar{X}=R, \bar{Z}(\omega_3)=F\rangle$ \\ \hline
    SU(4), $\mathbf{4}$ &   II & 4 &    \includegraphics[width=.1\textwidth]{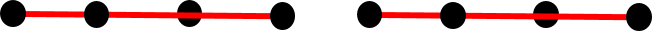}
    & $P^{\{5\}}_{(4,\bar{20})}$  & 4-level, $\langle \bar{X}=T, \bar{Z}=F\rangle$ \\ \hline
    SU(4), $\mathbf{6}$ &    I & 2 &   \includegraphics[width=.1\textwidth]{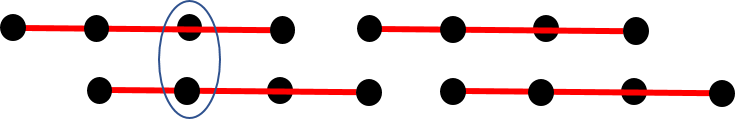}
    & $P^{\{5\}}_{(20\otimes\bar{20})}$  & 4-level, $\langle \bar{X}=T, \bar{Z}=F\rangle$ \\ \hline
    SU(4), $\mathbf{10}$ & II & 2 &  \includegraphics[width=.1\textwidth]{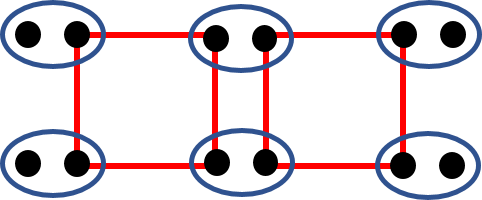}
    & $P^{\{2\}}_{(20',45)}$  & unique \\ \hline
    SU(4), $\mathbf{15}$ &  I & 1 &   \includegraphics[width=.1\textwidth]{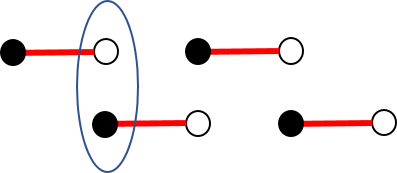}
    & $P^{\{2\}}_{(\bar{4}\otimes 4)}$  & qubit, $\langle \bar{X}=\Pi, \bar{Z}(\omega_4)=F\rangle$ \\ \hline
    SU(4), $\mathbf{20'}$ & I & 1 &   \includegraphics[width=.08\textwidth]{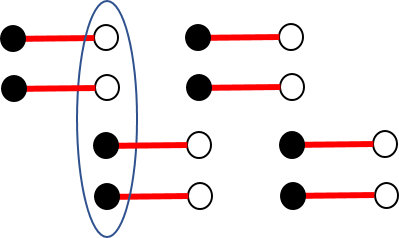}
   & $P^{\{2\}}_{(\bar{4}\otimes 4 \otimes \bar{4}\otimes 4)}$  & qubit, $\langle \bar{X}=\Pi,
    \bar{Z}(\omega_4)=F\rangle$ \\ \hline
    SU(4), $\mathbf{35}$ & II & 1 &   \includegraphics[width=.1\textwidth]{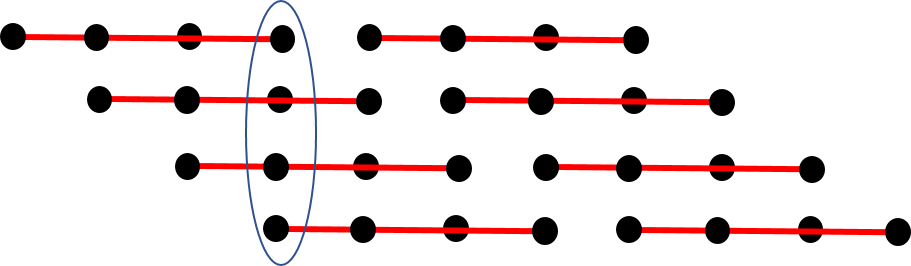}
    & $P^{\{2\}}_{(4\otimes 4 \otimes 6\otimes 6\otimes 6)}$  & unique \\ \hline
    SU(4), ($\mathbf{4}$, $\mathbf{\bar{4}}$) & III & 2 &   \includegraphics[width=.1\textwidth]{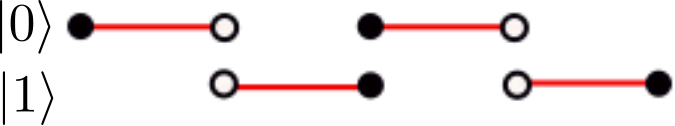}
    & $P^{\{3,A\}}_{(4)}$, $P^{\{3,B\}}_{(\bar{4})}$  & qubit, $\langle \bar{X}=R, \bar{Z}(\omega_4)=F\rangle$ \\ \hline
    SU(4), ($\mathbf{10}$, $\mathbf{\bar{10}}$) & III & 2 &    \includegraphics[width=.08\textwidth]{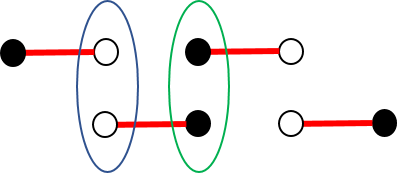}
    & $P^{\{2\}}_{(\bar{4}\otimes 4)}$  & unique \\ \hline \hline
  \end{tabular}
\end{table*}

In this section, we study qubits using various 1D SU($N$) VBS models~\cite{AKLT87}.
Difference from SU(2) case, there are more degree of freedoms and,
as a result, non-standard (i.e., non-Clifford~\cite{NC00}) logical operations exist naturally,
and we find there are three types of encodings,
see Table~\ref{tab:vbsqs}.

For SU(2), irreps are classified into two classes:
integer spins as linear reps of $SO(3)=SU(2)/\mathbb{Z}_2$,
half-integer spins as projective reps of SO(3).
Tensor product of two half-integer spins leads to a direct sum of integer spins.
For SU($N$), irreps are classified into $N$ classes according to
their congruence classes~\cite{Ram10}.
Let $[\lambda]$ be the congruence class of a rep $\lambda$,
then $[\lambda]\in \{[0],[1],\dots,[N-1]\}$.
For example, for SU(3), irrep $\mathbf{3}$ belongs to the class [1],
irrep $\mathbf{\bar{3}}$ belongs to the class [2],
irrep $\mathbf{8}$ belongs to the class [0].
A simple way to find the congruence class of an irrep
is that it is the number of boxes in its Young tabular modular $N$.

The SU($N$) VBS states are constructed with linear reps of $PSU($N$)=SU($N$)/\mathbb{Z}_N$,
which then allows $N$ different SPT phases labelled by $\omega_N^{[\lambda]}$
for $[\lambda]$ as a label of the congruence class of its virtual reps.
If the on-site irrep is of class $[\lambda]$ which is not linear,
i.e., class $[0]$,
then we can group $r$ nearest sites together such that $\omega_N^{r[\lambda]}=1$
for the smallest integer $r$.
Furthermore, even when $[\lambda]$ is a linear rep,
it can be complex, such as the rep $\mathbf{10}$ of $SU(3)$,
so a SU($N$) VBS system can be non-translation-invariant by one lattice site.

The algorithm to construct a SU($N$) VBS system,
including its ground states and parent Hamiltonian,
is as follows.
For translation-invariant (TI) system, given the on-site irrep $\lambda$, find the minimal integers $x$ and $y$ as
the number of virtual $\mathbf{N}$ and $\mathbf{\bar{N}}$, respectively,
the product of which leads to $\lambda$.
Next form singlet bond from $N$ irreps $\mathbf{N}$ (or $N$ irreps $\mathbf{\bar{N}}$ ),
then a VBS state is constructed.
In terms of MPS, the set of matrices for each site follows from Clebsch–Gordan (CG) coefficient
of $x\cdot \mathbf{N} \otimes y \cdot \mathbf{\bar{N}} \rightarrow \lambda$.
The SPT order is determined by its edges.
The parent Hamiltonian contains at most $(N+1)$-local terms.

For non-TI (NTI) system, there are two sublattices $\textsc{a}$ and $\textsc{b}$.
Given the on-site irrep $\lambda$ and $\bar{\lambda}$, find the minimal integers $x$ and $y$ as
the number of virtual $\mathbf{N}$ and $\mathbf{\bar{N}}$, respectively,
the product of which leads to $\lambda$.
Next form singlet bond from $\mathbf{N}$ and $\mathbf{\bar{N}}$,
then a VBS state is constructed.
In terms of MPS, the set of matrices for each $\textsc{a}$ ($\textsc{b}$) site follows from CG coefficient
of $x\cdot \mathbf{N} \otimes y \cdot \mathbf{\bar{N}} \rightarrow \lambda$
($y\cdot \mathbf{N} \otimes x \cdot \mathbf{\bar{N}} \rightarrow \bar{\lambda}$).
The SPT order is determined by its edges.
The parent Hamiltonian contains at most three-local terms.

For an irrep of SU($N$) with m boxes in its Young diagram,
its generators can be expressed with the set of $N^2$ operators
\begin{equation}\label{}
  S^\alpha_\beta := \psi^{\dagger \alpha c} \psi_{\beta c} -\delta^\alpha_\beta \frac{m}{N}
\end{equation}
for fermionic operator $ \psi^{\dagger \alpha c}$ as the creation operator of a state with flavor $\alpha$
and color $c$ and $\{\psi^{\dagger \alpha a},  \psi_{\beta b}\}=\delta^\alpha_\beta \delta^a_b$~\cite{Aff85}.
Their possible values are $\alpha=1,2,\dots,N$, and $c=1,2,\dots,p$ for $p$ as the number of columns
in its Young diagram.
Einstein's summation role is assumed.
The total number of boxes is $m=\psi^{\dagger \alpha c} \psi_{\alpha c}$,
and the set $S^\alpha_\beta $ satisfies $S^\alpha_\alpha=0$.

For SU($N$) irreps, as there can be multiple rows,
we can also use `hole' operator $\bar{\psi}^\dagger_\beta:=\psi_\beta$,
which corresponds to a column of $N-1$ boxes in Young diagram.
For the fundamental irrep $\mathbf{N}$, we have
$ S^\alpha_\beta = \psi^{\dagger \alpha} \psi_{\beta} -\delta^\alpha_\beta \frac{1}{N}$,
and $\psi^{\dagger \alpha} \psi_{\alpha}=1$.
For its conjugate $\mathbf{\bar{N}}$, we have
$ S^\alpha_\beta = \psi^{\dagger \alpha} \psi_{\beta} -\delta^\alpha_\beta \frac{N-1}{N}$,
and $\psi^{\dagger \alpha} \psi_{\alpha}=N-1$.
Using hole operator, it can also be expressed as
$ S^\alpha_\beta =- \bar{\psi}^{\dagger}_\beta \bar{\psi}^{\alpha} +\delta^\alpha_\beta \frac{1}{N}$,
and $\bar{\psi}^{\dagger }_\alpha \bar{\psi}^{\alpha}=1$.
For the adjoint irrep $\mathbf{N^2-1}$, we have
$ S^\alpha_\beta =\psi^{\dagger \alpha} \psi_{\beta}  - \bar{\psi}^{\dagger}_\beta \bar{\psi}^{\alpha}$,
which is the combination of a particle and a hole.

For the twist operator,
we first notice that there exists an operator $\mathcal{O}_\lambda$ in the Cartan subalgebra such that
\begin{equation}\label{}
  e^{i2\pi \mathcal{O}_\lambda}=\omega_N^{[\lambda]}\mathds{1},\; \omega_N=e^{i2\pi/N},
\end{equation}
as the analog of that of SU(2) case.
For the SU(3) example, we can choose $\mathcal{O}=T^3+\frac{1}{\sqrt{3}}T^8$
and it is easy to see that it satisfies the relation above.
The choice of $\mathcal{O}_\lambda$ is not unique; however,
as we can see below with the MPS picture or the fermion representation,
there exists a simple way to choose it.

Now we first study the twist for SU(2) VBS.
For SU(2) irreps, there is only one row in its Young diagram,
and for $p$ boxes, we have $p=2S$ for $S$ the spin value.
As there are only two flavors, we let $ a^{\dagger c}:=\psi^{\dagger 1 c}$,
$ b^{\dagger c}:=\psi^{\dagger 2 c}$, $c=1,2,\dots,p$.
The z-component of spin is $S^z=(a^{\dagger c} a_c- b^{\dagger c}b_c)/2$,
the ladder operator is $S^+=a^{\dagger c} b_c$.
Using fermion operators, a VBS defined in Eq.~(\ref{eq:vbssu2}) can be expressed as
\begin{equation}\label{}
  |\Xi_{mn}\rangle =\prod_{c=1}^n \prod_{r\in odd} B_{c,r}
 \prod_{c=1}^m  \prod_{r\in even} B_{c,r} |\Omega\rangle
\end{equation}
for $|\Omega\rangle$ as vacuum state,
$B_{c,r}:=a^{\dagger c}_{r} a_{c,r+1}+b^{\dagger c}_{r} b_{c,r+1}$.
The action of twist $F$ on site $n$ is
\begin{equation}\label{}
a^{\dagger c}_{r} \rightarrow e^{i\ell r/2} a^{\dagger c}_{r}, \;
b^{\dagger c}_{r} \rightarrow e^{i\ell r/2} b^{\dagger c}_{r}.
\end{equation}
Recall that $\ell:=2\pi/L$.
Now the twist depends on whether the last site has $n$ or $m$ bonds to its right.
Then we find $\langle \Xi_{mn}|F|\Xi_{mn}\rangle=(-1)^n$ in the large-$L$ limit.
Note that the action of twist can be equivalently treated as
$a^{\dagger c}_{r} \rightarrow a^{\dagger c}_{r},
b^{\dagger c}_{r} \rightarrow e^{i\ell r} b^{\dagger c}_{r}$,
i.e., only one flavor is affected,
but with an additional global phase accumulated from each bond,
which can be absent by modifying the form of twist,
e.g., change $S^z_r$ to $S^z_r-\mathds{1}/2$ or others.
As the operator $S^z_r$ is natural for SU(2) case,
we will use it for the twist which is suitable for implementation by external field,
while for SU($N$) case we also use twist that only affects one flavor.

Now, for SU($N$) VBS, we distinguish three types:
I) TI system with on-site real irrep.
The GSD comes from SSB of parity (reflection about a link).
II) TI system with on-site complex irrep.
The GSD comes from SSB of lattice translation.
III) NTI system with complex irrep $\lambda$ on odd sites and $\bar{\lambda}$ on even sites.
The GSD comes from SSB of parity (reflection about a site).

The singlet formed by $\mathbf{N}$ and $ \mathbf{\bar{N}}$ from two nearest neighbouring sites is
\begin{equation}\label{}
  |\omega\rangle = \frac{1}{\sqrt{N}} \sum_{i=0}^{N-1} |i\bar{i}\rangle
\end{equation}
for $\{|i\rangle\}$ as a basis of $\mathbf{N}$ and
$\{|\bar{i}\rangle\}$ as the corresponding basis of $ \mathbf{\bar{N}}$.
The singlet formed by $N$ product of irreps $\mathbf{N}$ from neighbouring sites,
which is a so-called n-mer, or extended valence bond~\cite{GR07}, is
\begin{equation}\label{}
  |s\rangle = \frac{1}{\sqrt{N}} \sum_{i_1,i_2,\dots, i_N}^{N-1} \varepsilon_{i_1,i_2,\dots, i_N} |i_1,i_2,\dots, i_N\rangle
\end{equation}
for symmetric tensor $\varepsilon_{i_1,i_2,\dots, i_N}$, known as Levi-Civita symbol.
An SU($N$) VBS can be expressed as a product of singlet $|\omega\rangle$ or $|s\rangle$.

For type-I, suppose $\lambda$ is from a minimal of $\eta$ product of $\mathbf{N}$ and $\mathbf{\bar{N}}$.
Then there are in total $\eta L$ bonds.
A bond can be a \emph{left} (\emph{right}) bond if $\mathbf{N}$ is to the left (right) of $\mathbf{\bar{N}}$.
Let the number of left and right bonds be $\eta_L L$ and $\eta_R L$, then
$\eta=\eta_L+\eta_R$.
Now a VBS is a product of $\eta_L L$ left bonds and $\eta_R L$ right bonds, denoted as $|\Xi_{\eta_L\eta_R}\rangle$.
The twist operator only induces a phase factor for one flavor, say, $i$, with
\begin{equation}\label{}
 \bar{\psi}^{\dagger}_i  \psi^{\dagger i} \rightarrow e^{i\ell} \bar{\psi}^{\dagger}_i  \psi^{\dagger i},
\end{equation}
and then a phase $e^{-i\ell/N}$ on a left bond, $e^{i\ell/N}$ on a right bond.
Then
\begin{equation}\label{}
  \langle \Xi_{\eta_L\eta_R}| F |\Xi_{\eta_L\eta_R}\rangle =e^{i2\pi (\eta_R-\eta_L)/N}.
\end{equation}
For instance, for the two degenerate ground states of SU($N$) VBS with on-site adjoint irrep~\cite{WAR18},
we get the phases $e^{\pm i2\pi/N}$ as $\eta_R-\eta_L=\pm 1$,
and for its fully dimerized excited states, we get the trivial phase as $\eta_R=\eta_L$.

For type-II, the on-site complex irrep $\lambda$ can be from a product of $\mathbf{N}$,
denote the minimal number as $\eta$.
As a result, there are only irreps $\mathbf{N}$ as the virtual particles,
and they can form a n-mer.
To study the twist effect,
first notice that there will be a factor $\omega_N^{[\lambda]}$ on the last site.
The twist effect will be similar to that on SU(2) VBS, namely,
the nontrivial phase depends on the edge structure,
and there is no phase accumulation from the bonds in the bulk in the large-$L$ limit.
As there are $\eta$ n-mer attached to the last site,
the formula of the nontrivial phase is more complicated than the SU(2) case;
however, it can be easily computed case by case.
We can see this from our examples.

For type-III,  if irrep $\lambda$ is from a product of minimal $\eta$ irreps $\mathbf{N}$,
then $\bar{\lambda}$ is from  a product of minimal $\eta$ irreps $\mathbf{\bar{N}}$.
The twist effect will be a combination of the effects from type-I and type-II models.
Namely, suppose $\eta=\eta_L+\eta_R$, then there will be a factor
$e^{i2\pi (\eta_R-\eta_L)/N}$ from the bonds in the bulk,
and there is also a factor $\omega_N^x$ from the last site,
for $x$ as a certain function of $[\lambda]$.
Notice that for the twist on NTI system, it takes the form
\begin{equation}\label{}
  F(\{\theta_n\},\theta)=\otimes_{n=1}^L e^{i(-1)^n \theta_n g_n}.
\end{equation}
The reason is that the even and odd sites are the conjugate of each other.
We present examples for SU(3) and SU(4) cases in Table~\ref{tab:vbsqs}.
The GSD can be seen from the edge structure of a unit cell.
Note we use $P_{(\lambda)}$ as the projector onto all other irreps except $\lambda$,
which can be easily understood from the construction of VBS models.
Many VBS models have been constructed before~\cite{Aff85,GR07,Aro08}.

We call a unit cell as the grouped site such that it is a linear rep.
For encoding, we find the following general features.
For TI systems that break translation, it encodes $N$-level logical system,
$\bar{X}=T$, $\bar{Z}=F$.
For NTI systems that break translation (or parity), it encodes a logical qubit, $\bar{X}=T$ (or $\Pi$), $\bar{Z}(\omega_N)$ is twist.
Here $\Pi$ denotes the generator of parity symmetry.
We emphasize that, for NTI system we obtain non-Pauli logical operators $\bar{Z}(\omega_N)$.
This feature is not present for SU(2) VBS qubits.

For practical implementation of the twist, it seems harder than the SU(2) case, though.
However, based on ideas from quantum simulators~\cite{BN09},
if SU($N$) VBS can be realized in well-controlled artificial systems,
such as superconducting devices, optical lattices, or trapped ions,
a global twist can also be simulated by a certain flux insertion process.
This warrants a separate study.

\section{2D TOP and SPT qubits}
\label{sec:2D}

In this section we study and compare some classes of 2D topological qubits briefly.
Distinct from the 1D case,
there are more orders for 2D many-body systems,
especially TOP order.
Note that TOP order can be roughly viewed as a symmetry-breaking of high-form symmetry~\cite{KS14,GKS+15,Wen19},
and SSB refers to symmetry-breaking of a global symmetry.
We find SET qubits are novel and there is a nontrivial interplay
between Wilson loops and global twists.
We also compare the classes of SPT and SET qubits,
and point out their connections.
We only highlight the features that are relevant for this work.

\subsection{2D SET qubits}

\begin{figure}
  \centering
  \includegraphics[width=.4\textwidth]{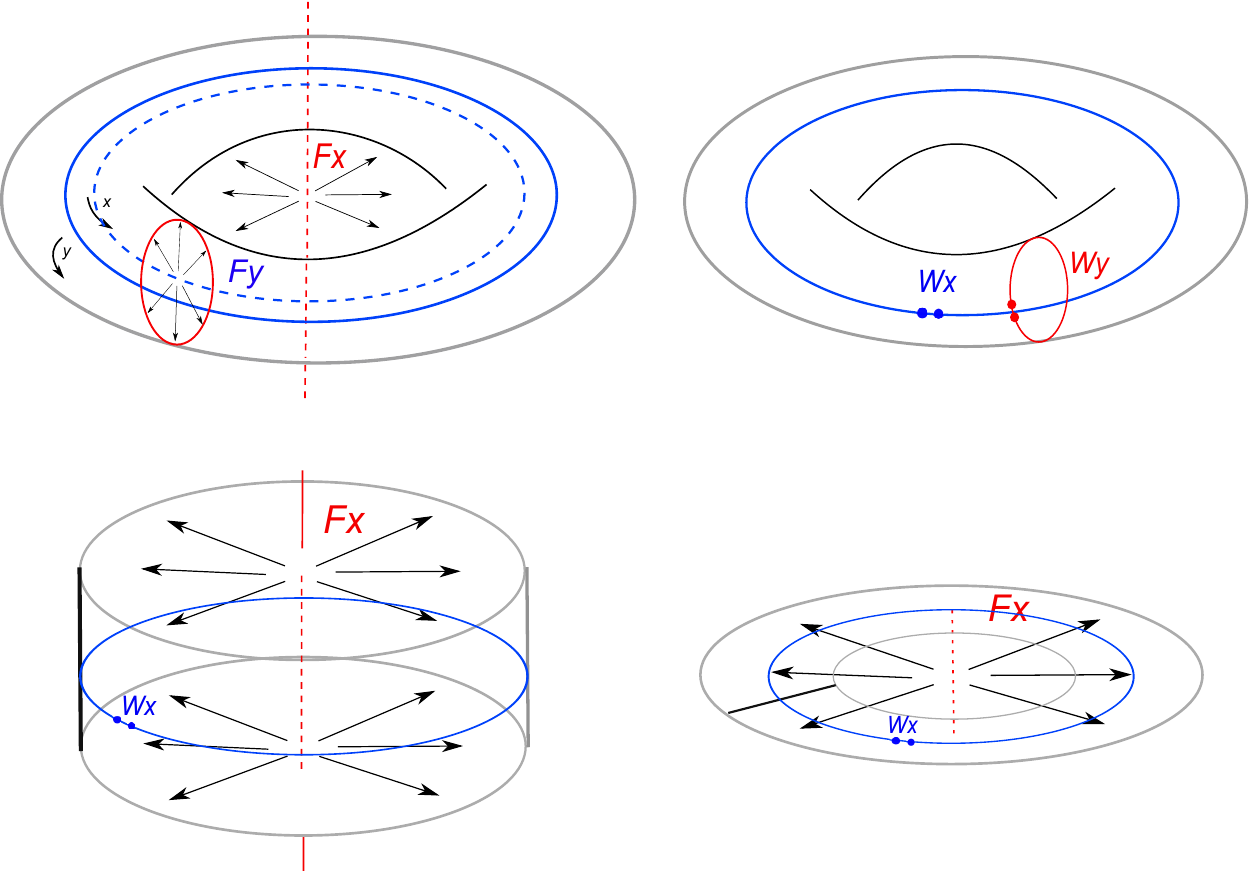}
  \caption{Equivalence between twist (flux insertion) and Wilson loop.
  Up-left: twists $[F_x, F_y]=0$; Solid blue and red circles are grouped sites for twists.
  Dashed blue and red lines are electric lines.
  Up-right: Wilson loops depend on statistics of excitations.
  Down-left: twist and Wilson loop on cylinder.
  The black line is grouped site.
  Down-right: twist and Wilson loop on disc.
  The black line is grouped site.}\label{fig:twistandwilson}
\end{figure}

In this subsection we first study 2D SET qubits with continuous global symmetry.
A unique feature in this setting is the relation between Wilson loop and twist operations,
see Fig.~\ref{fig:twistandwilson}.
In this work we only consider finite abelian gauge groups.
We find that logical operators can be played by both twist and Wilson loops.
We observe that they play different roles in the encoding.
Wilson loop determines the code distance,
while twist is more suitable for global implementation.
A twist is equivalent to a corresponding Wilson loop, while
Wilson loops are inevitable as they cannot be fully substituted by twists.
Code distances are the square root of the system size $d_x=d_z=\sqrt{L}$,
and it is symmetric for logical bit flip and phase flip operations.
As noncommuting Wilson loops intersect only on a few sites,
the logical support is a small constant independent of the system size.
This applies to the quantum dimer models (QDM)
and other resonating valence-bond (RVB) models~\cite{And87,RK88,Bon89,MS01,MSP02,LYC+14,MOP16}.

Note that VBS and RVB states can be put in the same framework of tensor-network states.
There can be SET order if the virtual irreps belong to different projective classes~\cite{SPC+12},
and there exists SPT order if the virtual irreps belong to the same projective class.
A RVB state can be expressed as an equal-weight superposition of VBS states.
A ground state with SET order is a RVB state,
and with SPT order is a VBS state.
Below we use QDM and fractional quantum hall (FQH) states as examples
to highlight their features.

FQH with filling factor $\nu=1/q$ ($q\in 2\mathbb{Z}+1$) on a torus has ground state degeneracy $q$.
FQH is the example that the two flux operators $F_x$ and $F_y$ do not commute.
Note we only consider Laughlin abelian case, the non-abelian cases are more involved~\cite{WN90}.
FQH is U(1)-SET phase as the electron number is conserved.
Due to the global U(1) symmetry, we can use twist (inserting flux).
There is only one species of excitation: chargon with fractional charge $e/q$, so only one type of flux.
We define the two directions of torus as x and y.
The GSD encodes a q-level system,
and there are many ways to define its logical operators
\begin{equation}\label{}
  \bar{X}:=T_x=F_y=W_x,  \bar{Z}:=T_y=F_x=W_y,\; \bar{X}^q=\bar{Z}^q=\mathds{1}.
\end{equation}
$T$ is magnetic translation operator~\cite{Wen90},
$W$ is Wilson loop (create quasipartile-hole pair,
then transport one along a nontrivial loop, then annihilate),
$F$ is twist.
$W_x$ is equivalent to inserting a flux $F_y$,
and translation $T_x$ is equivalent to $W_x$.
For the implementation issue, the flux operators are relatively easier,
given that we do not have clear clue how to implement $T$ and $W$ easily.

On the contrary, the well-known $Z_2$ gauge model,
e.g., the QDM and the toric code, is an example
that the two flux operators $F_x$ and $F_y$ commute.
For $Z_2$ gauge model there are two species of excitations: spinon and vison.
So there are two types of fluxes: electric and magnetic.
The spinon and vison are both self-boson,
while spinon and vison mutually is semion.
The spinon has fractional magnetic moment and shows symmetry fractionalization
with respect to $SO(3)$ symmetry,
so we can insert electric flux
by an analog of AC effect for the 1D case.

Denote spinon as $e$, vison as $m$.
Let Wilson loops on a torus be $W_{x,e}$, $W_{x,m}$, $W_{y,e}$, $W_{y,m}$, then
\begin{equation}\label{}
[W_{x,e},  W_{x,m}]=[W_{x,e},  W_{y,e}]=[W_{x,m},  W_{y,m}]=[W_{y,e},  W_{y,m}]=0,
\end{equation}
and
\begin{equation}\label{}
\{W_{x,e},  W_{y,m} \}=\{W_{x,m},  W_{y,e} \}=0.
\end{equation}
As spinon $e$ is fractionalized, there are twists $F_{x,e}$ and $F_{y,e}$ such that
\begin{equation}\label{}
\{F_{x,e}, W_{x,e}\}= \{F_{y,e}, W_{y,e}\}=0.
\end{equation}
So
\begin{equation}\label{}
F_{x,e}=W_{y,m}, F_{y,e}=W_{x,m}, [F_{x,e}, F_{y,e}]=0.
\end{equation}
The two twists (flux insertion) commute but not with Wilson loop.

A simple understanding of the independence of $F_{x,e}$ and $F_{y,e}$ is from the dimer covering.
In quantum dimer model,
a ground state is an equal-weight superposition of short-range dimer covering of the lattice.
The local $Z_2$ gauge condition converts to even/odd parity for the number of bonds along x and y cut.
The dimer covering breaks translation symmetry, like the 1D MG model.
The twist for each direction only induces logical action for one direction, x or y, not both,
as a bond orthogonal to the twist direction do not accumulate phase factors.
In addition, we note that this can also be captured by Chern-Simons field theory~\cite{WN90}
and the so-called BF field theory.

Furthermore, there is a dependence on the size of the system~\cite{Bon00}.
As each dimer is formed by two spins,
the system should have even number of sites $L=L_xL_y \in 2\mathbb{Z}$.
On a cylinder, the GSD is two.
Given $L_x\in 2\mathbb{Z}$, $L_y$ can be even or odd.
For any $L_y$, logical $\bar{Z}$ operator is provided by twist $F_x$.
For odd $L_y$, logical $\bar{X}$ operator is provided by translation operator $T_x$ by odd number of lattice sites,
for even $L_y$, logical $\bar{X}$ operator is a Wilson loop,
which requires breaking dimers and drag soliton around the system and come back to annihilate again.
This is similar to the 1D MG model.
On square lattice with $L_x=L_y=L$ and $L$ has to be even, $L\in 2\mathbb{Z}$,
there are four ground states instead of two.
The twist does not act as logical $\bar{Z}$ as the even number of sites in both directions
trivializes the twist phase.
Also translation operator does not act as logical $\bar{X}$.
The only logical operators in this case are Wilson loops.

Finally, we remark on a difference between SET qubits and standard TOP qubits.
By definition, TOP qubits do not have preserved global symmetry,
hence there will be no twist or flux insertion.
TOP qubits are the most well understood class of qubits,
such as TOP stabilizer codes and subsystem codes~\cite{BT09,BPT10,BK13,LP13,KLP+05,KLP+06}.
Limitations on code distance have been established~\cite{BT09,BPT10,BK13},
while here we also highlight the limitation on logical support.

Take the well-known toric code as the example,
which is also known as $Z_2$ gauge model.
A usual phase diagram~\cite{KL09,TKP+10} is depicted in Fig.~\ref{fig:phases}(Right).
The gapped TOP phase is deconfining, although there is a slight confining force among anyons.
The GSD can be understood as a SSB of symmetries defined by Wilson loops,
$X_\ell$ or $Z_\ell$ for any homologically nontrivial loop $\ell$,
which is a 1-form symmetry.
The algebra of Wilson loops defines anyon braiding statistics.
On a torus two qubits are encoded,
with $\bar{X}_x=X_{\ell_x}$, $\bar{Z}_x=Z_{\ell_y}$,
for the first qubit,
$\bar{X}_y=X_{\ell_y}$, $\bar{Z}_y=Z_{\ell_x}$,
for the second qubit,
and x, y as the two directions of the torus.
The code distance is $\sqrt{L}$ as the minimal weights for $\bar{X}$ and $\bar{Z}$
are both $\sqrt{L}$.
The logical support is 1,
similar to the classical 2D Ising model.
The low support might make the qubit vulnerable to local errors on the intersection site of two Wilson loops,
yet, on the other hand,
it could benefit local stabilizer measurements
and active error correction when these need to be performed.

The small logical support can be attributed to
the connection between toric code and Ising model (with $XX$ terms).
The toric code can be viewed as a coupled system of 1D Ising wires,
$Z_{\ell_x}$ and $Z_{\ell_y}$
are the symmetry being broken by 1D Ising wires,
and $X_{\ell_x}$ and $X_{\ell_y}$
are symmetry being preserved.
Hence TOP order can be understood as a combination of SPT of a 1-form symmetry (by loop of $X$)
and SSB of a 1-form symmetry (by loop of $Z$)~\cite{Wen19}.
Other (abelian) topological subsystem stabilizer codes~\cite{KLP+05,KLP+06} have similar properties with toric code.
For instance, the 2D Bacon-Shor code~\cite{Bac06},
which belongs to the class of compass model~\cite{NB15} and might be gapless~\cite{DBM05},
is defined by a gauge group,
has logical shape as string, distance $\sqrt{L}$,
and logical support 1.

\subsection{SPT vs. SET qubits}

Here we compare SPT and SET qubits in a bit more details,
both of which benefit from the preserved symmetry.
Although the SPT qubits, especially VBS qubits, are defined for 1D systems,
and SET qubits are for 2D systems,
we find the dimensionality does not play central roles here.

To define 2D SPT qubits,
the class of 2D VBS models with global SU($N$) symmetry are the natural physical systems.
We find the features of encoding carry over from the 1D case,
so it is not necessary to generalize SPT qubits from 1D to 2D
for the purpose of better encoding.
However, for completeness, we highlight some new features.
Compared with 1D case, the complexity of 2D systems is that
their properties depend on the underlying geometry of the lattices.
To determine whether a model is gapped or gapless in general is a nontrivial problem.
Our encoding only works for the case when there is a gap.

To define a twist for 2D case, $n$ is a single grouped-site of all sites along
the other direction with the same site index along the periodic direction.
For 2D system on a torus, we can define two different twists,
corresponding to flux insertion for the two `holes'.
Sites in each row or column need to be grouped together and treated as a single site,
and the system size entering the twist operator is $L_x$ or $L_y$ for the two directions x or y.
The twist phase is also easy to compute using the fermion representation.
It generally includes two part: one part from the action on the edge (the last blocked site),
and the other from the action in the bulk.
For instance, the SU(3) simplex solid with on-site $\mathbf{3}$ irrep on Kagom\'{e} lattice~\cite{Aro08}
shows a SSB of lattice symmetry, hence can encode a qubit,
as an analog of the 1D MG model.

As passive quantum memory,
systems with SET or SPT order would behave similarly as topological qubits,
since for both of which
the broken or preserved symmetries are nonlocal.
We have to consider nontrivial quantum gates to resolve their differences and connections further.
Due to TOP order, SET qubits can also be encoded into anyonic excitations
(or holes, defects)~\cite{BLK+17},
braiding of which can enable universal quantum computing.
While SPT order, by definition, does not support anyons;
however, these systems also enable universal quantum computing~\cite{GE07,WAR11,Miy11,WSR17}.
A promising approach is to use `extrinsic' defects in SPT systems,
which, so far, have not been fully understood~\cite{BBC19}
except for a few notable systems~\cite{MCA+14,AHM+16,PCG+12}.
Coupling of 1D VBS wires, with junctions as a sort of defects,
has the potential for universal quantum computing~\cite{Wang19}.
Such networks of SPT qubits are effectively 2D systems with a large number of defects,
similar with the anyonic encodings in TOP or SET systems.
In addition, compared with 2D systems,
different 1D SPT qubits can be properly wired up in principle,
hence their logical gates can be combined together leading to higher computational power.

\section{Conclusion and Discussion}
\label{sec:conc}

In this work, we mainly studied the class of 1D SPT qubits
and compared with other classes of topological qubits on a `logical' level,
with their main features summarized in Table~\ref{tab:qubits}.
When a global U(1) symmetry is present and preserved,
a topological twist operator exists and plays the role of a logical operator.
The class of SU($N$) VBS models can provide non-stabilizer codes
and gates that are on high levels of the Clifford Hierarchy~\cite{NC00}.
When there is only discrete global symmetry,
our study of Pauli Hamiltonian models, mainly the Wen model,
shows that flux insertion also exists and plays the role of a logical operator,
although there is no twist operator.
Beyond SSB order, for 2D systems there are TOP orders and
there is a nontrivial interplay between Wilson loops and twist for SET qubits
with a global U(1) symmetry.
Wilson loops determine the code distance,
while twist can increase the logical support and benefit practical global implementations.
Also, we shall note that when qubits are realized in real materials or artificial simulators,
many other factors will affect the characters and controllability of a qubit.

Along the line of research in this work,
further investigations can be taken in the future.
For instance,
2D SU($N$) SPT and SET qubits can be constructed,
which, however, require proofs of the existence of gap.
2D Pauli Hamiltonian models with SPT orders,
for both weak and strong SPT cases also exist in principle.
Also 3D topological qubits are highly nontrivial while important,
which may provide a thermally stable (self-correcting)  quantum memory~\cite{BT09,BPT10,CLBT10},
such as the fracton orders~\cite{VHF15}.
Comparison with the qubits encoded by defects or edge modes is also important.
Last but not least,
we did not study active quantum error correction and computation on them,
such as measurement, readout, and entangling gates,
which are inevitable but highly nontrivial.

\section{Acknowledgement}

This work has been funded by CIFAR and NSERC.
The author acknowledges Stewart Blusson Quantum Matter
Institute and Department of Physics and Astronomy,
University of British Columbia where the main
part of this work was performed,
I. Affleck and R. Raussendorf for thoughtful suggestions on the manuscript,
and S. Jiang, O. Kabernik, S.-P. Kou, G. A. Sawatzky, H. Tasaki, H.-H. Tu, Q.-R. Wang for discussions.

\appendix

\section{SPT qubits with discrete symmetries}
\label{sec:app}

Here we define examples of SPT qubits with discrete symmetries.
These examples are variations of cluster states and toric code,
which are notable examples with SPT order and TOP order, respectively.
They can also be viewed as generalizations of the five-qubit code~\cite{LMP+96}
to the setting of SPT orders,
which is the smallest code that can correct an arbitrary error.

\subsection{1D SPT qubits}
\label{sec:1ddis}

Now we turn to 1D SPT order by discrete symmetry.
As there is no $U(1)$ global symmetry, a twist cannot be directly defined.
However, flux insertion is still allowed which will induce geometric phase.
The flux insertion process can be treated as a discrete version of the twist for $U(1)$ case.
To define a SPT qubit,
we look for models with discrete symmetry $G_1\times G_2$,
while there is SSB of $G_1$ and SPT of $G_2$.
Below we study a class of Pauli Hamiltonian model we term as `Wen model',
as the basic interaction term $XYYX$ firstly appeared in Wen's
model on the 2D square lattice~\cite{Wen03}.
It is also a generalization of the five-qubit code~\cite{LMP+96}.
For the 1D case, we find there is SPT order,
hence there is a GSD
to encode one logical qubit.
As the model is commuting, the large system size limit is not required.

We study 1D Wen model of $L$ two-level systems (qubits) with PBC
\begin{equation}\label{}
  H=-\sum_n X_{n-1} Y_n Y_{n+1} X_{n+2},
\end{equation}
each term in it is from a product of $X_{n-1} Z_n X_{n+1} $ and $X_{n} Z_{n+1} X_{n+2}$,
which takes the form of the stabilizers of the 1D cluster state~\cite{BR01}.
This model has also been considered in other settings such as phase transition~\cite{VMP17,FRR+17}.
The model can also be expressed as
\begin{equation}\label{}
  H=-\sum ( \nabla\mkern-9mu\triangle + \triangle\mkern-9mu\nabla ).
\end{equation}
As such the system lives on a zig-zag ladder.
The $YY$ term is on the diagonal of each diamond.
The total number of sites can be even or odd.
For odd number of sites, the lattice has to be geometrically twisted once
to satisfy periodic boundary condition.

The model is commuting, hence exactly solvable.
The model has two degenerate ground state due to SSB of a global $Z_2$ symmetry,
one as cluster state $|C_0\rangle=\bigotimes_n H_n \bigotimes_n CZ_n (|+\rangle)^{\otimes L}$ for $CZ_n$
as controlled-Z gate on sites $n$ and $n+1$, $H_n$ as Hadamard gate on site $n$,
and the other as  $|C_1\rangle=\bigotimes_n X_n |C_0\rangle$.
The logical space is spanned by $|C_0\rangle$ and $|C_1\rangle$.
As a result, logical $\bar{X}$ is the generator $\vec{X}:=\bigotimes_n X_n $ of the broken symmetry.
For odd number of sites, the logical $\bar{Z}=\bigotimes_n Z_n$
can be viewed as flux insertion operation.
The twist phase is $\pi$ for $|C_0\rangle$  and 0 for  $|C_1\rangle$.
This code includes the five-qubit code as a special case when $L=5$~\cite{LMP+96},
hence this model can also be viewed as a generalization of the standard five-qubit code.
Furthermore, $\bar{Z}$ can take other forms.
It is easy to see $X_{n-1} Z_n X_{n+1} $ can also serve as $\bar{Z}$,
as it anti-commutes with $\bar{X}$.
Also the weight of $\bar{X}$ can be reduced by about $2/3$ factor.
So we find the code distances $d_x\approx L/3$ and $d_z=3$.

\begin{table}[t!]
  \centering
\footnotesize{
 \begin{tabular}{|c|c|c|c|}
    \hline \hline
                  & 1D Ising & 1D Cluster & 1D Wen \\ \hline
    H term & $XX$ & $XZX$ & $XYYX$ \\ \hline
    SSB & $Z_2$ & no & $Z_2$ \\ \hline
    SPT & no & $Z_2\times Z_2$ & $Z_2\times Z_2$ \\ \hline
    electric charge  & no & monopole by $X$ & pair by $X$ \\ \hline
    magnetic charge   & pair by $Z$ & pair by $Z$ & pair by $Y$ \\ \hline
     logical qubit & $\bar{X}=\vec{Z}$, $\bar{Z}=X_n$ & no &  $\bar{X}=\vec{X}$, $\bar{Z}=\vec{Z}$ \\
    \hline \hline
  \end{tabular}}
  \caption{Comparison of features of 1D Ising model, cluster model, and Wen model.
  }\label{tab:1D}
\end{table}

Now we generalize the model to qudit case.
As there is a nontrivial connection with the cluster state,
we first define a qudit cluster model~\cite{WSR17}
\begin{equation}\label{}
  H=-\sum_n X^\dagger_{n-1} Z_n X^\dagger_{n+1} +h.c. ,
\end{equation}
and its unique ground state is $|C_0\rangle=\bigotimes_n F_n \bigotimes_n CZ_n (|+\rangle)^{\otimes L}$ for $CZ_n$
as qudit version of controlled-Z gate on sites $n$ and $n+1$,
$F_n$ as a Fourier operator on site $n$~\cite{WSR17}.
Several other cluster states are
 $|C_\ell\rangle=(X^{\ell})^{\otimes L}|C_0\rangle$ for $\ell=1,2,\dots,d-1$.
 To enforce the degeneracy of all $|C_\ell\rangle$, we define the 1D qudit Wen model
\begin{equation}\label{}
  H=-\sum_n X^\dagger_{n-1} Y_n^\dagger Y_{n+1} X^\dagger_{n+2} +h.c.,
\end{equation}
for $Y:=ZX$.
This includes the five-qudit code as a special case when $L=5$.
Also there is an asymmetry between the code distances of $\bar{X}$ and $\bar{Z}$.
This model has a global symmetry $Z_d\times Z_d\times Z_d$,
with the first two symmetry factors $Z_d\times Z_d$ for SPT order generated by
$ZIZ^\dagger IZI\cdots$ and $IZIZ^\dagger IZ\cdots$, respectively,
and the last one for SSB order generated by $\bigotimes_n X_n$.

The 1D Wen model is similar to the 1D Ising model $H_\text{Ising}=-\sum_n X_n X_{n+1}$.
Note we choose $XX$ term instead of $ZZ$ term for convenience.
There is a SSB of global $Z_2$ symmetry defined by $\vec{Z}:=\bigotimes_n Z_n$,
and a SPT of 1-local gauge symmetry defined by $X_n$ on any site $n$.
As $[H_\text{Ising},\vec{Z}]=[H_\text{Ising},X_n]=\{\vec{Z},X_n\}=0$,
the ground subspace can encode a qubit and serve as an error correction code,
with logical $\bar{X}=\vec{Z}$, $\bar{Z}=X_n$ for any $n$.
For odd number of sites on PBC, $\bar{Z}$ can also be $\bigotimes_n X_n$,
which is a flux insertion.
However, as a well known fact the minimal weight of $\bar{Z}$ is 1
(nontrivial action on a single site),
the Ising code is classical (also see section~\ref{sec:pre}).
In this regard, the 1D Wen model can be viewed as a quantum generalization of the 1D Ising model.

The 1D Ising model, cluster model, and Wen model can be related to each other.
By a `gauging' mechanism~\cite{Kog79,BBC19,LG12,Wang19Mb},
the cluster model $H=-\sum_i X_{i-1} Z_i X_{i+1}$ can be constructed
from the 1D Ising model by adding gauge qubit on each link and imposing the minimal coupling.
Roughly, gauging is to promote global symmetry to local gauge invariance,
which is neither unitary nor unique, and can change GSD in general.
The cluster state has SPT order from a global $Z_2\times Z_2$ symmetry,
with one factor from the Ising model,
and the other from the gauge qubits.
Then by folding the spectrum, the Wen model is obtained from the cluster model.
Their relations are summarized in Table~\ref{tab:1D}.

\subsection{2D SSPT qubits}

\begin{figure}
  \centering
  \includegraphics[width=.25\textwidth]{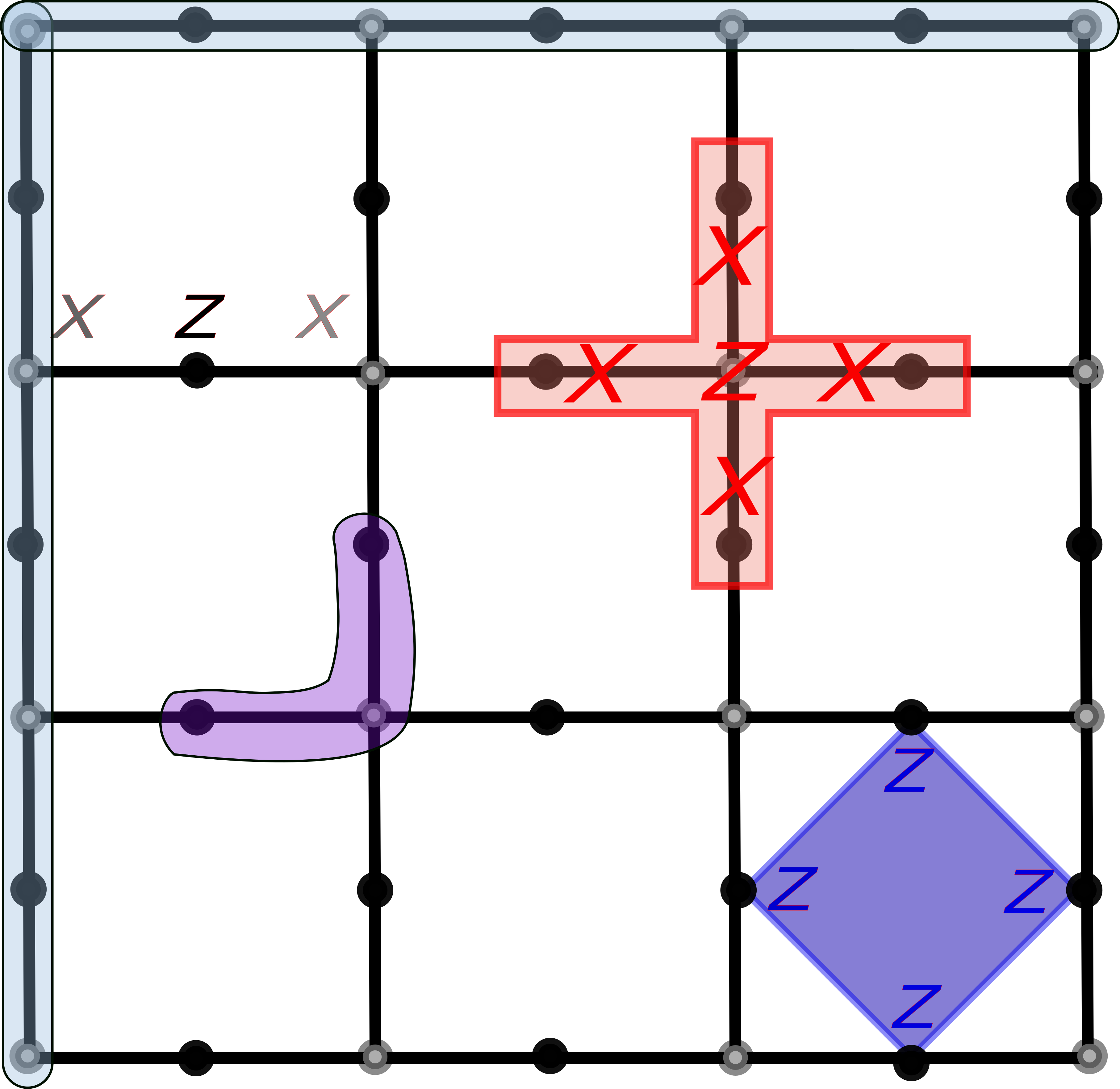}
  \caption{Schematic picture to show the procedure from 2D Ising model
  on the square lattice (gray dots) to the toric code.
  The Ising term $XX$ is modified to $XZX$ due to the addition of qubits on each link (black dots),
  and new terms $ZXXXX$ (shaded red loop) are added with $Z$ acting on each qubit in the original Ising model,
  resulting in a graph state as the network of vertical and horizontal 1D cluster states (shaded stripes).
  The unit cell of the graph state contains three qubits (shaded purple loop).
  The Z-stabilizer (shaded blue diamond) $ZZZZ$ of the toric code
  is obtained from product of four $XZX$ terms,
  and X-stabilizer $XXXX$ is obtained by measured out the original Ising system.
  }\label{fig:graph}
\end{figure}

As the final class of qubits that we consider,
we discuss an example with $Z_2\times Z_2$ symmetry,
while the first factor is for SSPT order,
and the second factor is for SSB order.
It can be generalized to the case $Z_d\times Z_d$ for integer $d>2$,
but we keep to the qubit case as the main features remain.
Also the study in this section can be viewed as a
2D generalization of that in section~\ref{sec:1ddis}.
The logical operators are from the order parameter directly as expected:
$\bar{X}$ is the generator of the broken symmetry,
and $\bar{Z}$ is the generator of the preserved symmetry.

We follow the procedure in section~\ref{sec:1ddis} that first involves
a gauging-like process and then folds spectrum to generate GSD.
Start from the 2D Ising model, we gauge it by adding qubits on each link.
Now the term $XX$ is changed to $XZX$,
and new terms $ZXXXX$ are added.
This is a graph state, as a coupled system of 1D cluster state wires,
see Fig.~\ref{fig:graph}.
The lattice can be treated as a square lattice with three qubits at each site.
Given the full set of stabilizers,
the graph state is the unique ground state.
The global $Z_2$ symmetry that is SSB by Ising model becomes a preserved subsystem symmetry.
Each wire in the graph state has $Z_2\times Z_2$ SPT order,
which is now a SSPT order of the whole graph state.
Next to induce GSD, we use spectrum folding as in the 1D case.
We replace the four weight-three stabilizers around a cell by their product $ZZZZ$,
then we obtain a model that is a `decorated' toric code.
The GSD is four on a torus.
The decoration changes the behavior of electric charges:
they can appear as monopole,
as the additional qubit in X-stabilizer cell serves as a sink of electric charge.
The decoration also affects the code distance:
the logical $\bar{Z}$ operator can be played by $XZX$ instead of a loop of $Z$,
so the code distance is three.

The toric code is obtained if the qubits in the original Ising model is measured out.
As mentioned, the toric code has SSPT order and SSB of a 1-form symmetry.
All logical operators are Wilson loops.
For a model to show SSPT and SSB orders,
it shall not have topological degeneracy and anyons.
We find this can be demonstrated by the Wen model on 2D square lattice~\cite{Wen03},
whose GSD has a nontrivial dependence of the system size.
Note that, by considering lattice symmetry,
the Wen model with different system sizes can be treated as different realizations
of Chern-Simons theory~\cite{KLW08}.
However, we do not consider point-group symmetry protection in our framework.

The 2D Wen model on square lattice is
\begin{equation}\label{}
  H=g\sum_{m,n} X_{m,n} Y_{m-1,n} Y_{m,n+1} X_{m-1,n+1},
\end{equation}
and we focus on the $g<0$ case.
This model can be generalized to the qudit case
as we did in section~\ref{sec:1ddis} for the 1D case,
while here we stick to the qubit case.
Let the size on the two periodic directions be $n$ and $m$, and total number of sites as $L=mn$.
On even by even square lattice,
its GSD is four,
and this case is equivalent to the toric code.
Let the two directions be the direction x and y.
$\bar{X}_x$ can be realized by $XYXY\cdots$ along direction x,
$\bar{X}_y$ can be realized by $XYXY\cdots$ along direction y,
$\bar{Z}_x$ can be realized by $Z_\ell$ along direction y,
$\bar{Z}_y$ can be realized by $Z_\ell$ along direction x.

On even by odd square lattice,
its GSD is two.
As the two directions are not equivalent,
anyons can only move freely along one of the two directions,
hence a reduction of the GSD from four to two.
Let the direction with even (odd) number of sites be the even (odd) direction.
$\bar{X}$ can be realized by $\bigotimes_n X_n$ on all sites,
or $XYXY\cdots$ along an even direction.
$\bar{Z}$ can be realized by $\bigotimes_n Z_n$ on all sites,
or $Z_\ell$ along an odd direction.
This means that a global operator is equivalent to a Wilson loop,
and logical operators of the encoded qubit come from Wilson loops.

On odd by odd square lattice,
its GSD is also two,
but it does not come from the usual TOP order.
The operator $XYXY\cdots$ cannot be defined along any of the two directions
due to the oddness of system size.
This means that excitations cannot be divided into two species,
only the fermions, as combined electric and magnetic charges, are left.
The GSD is due to SSB of global $Z_2$ symmetry,
and with the line symmetry by $Z_\ell$ it has SSPT order.
As is well studied, the line symmetry by $Z_\ell$ can be viewed as a flux insertion operation.
Therefore, we find that for the encoded qubit
$\bar{X}$ can be realized by $\bigotimes_n X_n$ on all sites,
$\bar{Z}$ can be realized by $Z_\ell$ on a line.
If we treat it as the 2D generalization of the five-qubit code,
it is obvious to see that the code distance grows as $\sqrt{L}$, instead of a constant,
due to the emergence of the line symmetry by $Z_\ell$.

\end{spacing}

\bibliography{ext}{}
\bibliographystyle{unsrt}

\end{document}